\pdfoutput=1
\RequirePackage{ifpdf}
\documentclass{JINST}

\usepackage[utf8]{inputenc}
\usepackage{amsmath}
\usepackage{subfig}		
\usepackage{tabularx} 			
\usepackage{units}			
\usepackage{url}

\graphicspath{{pic/}}

\title{Track segments in hadronic showers in a highly granular scintillator-steel hadron calorimeter}

 \author{\centering 
\LARGE\bf The CALICE Collaboration
}

\author{\centering
C.\,Adloff, 
J.-J.\,Blaising, 
M.\,Chefdeville, 
C.\,Drancourt,
R.\,Gaglione, 
N.\,Geffroy, 
Y.\,Karyotakis, 
I.\,Koletsou, 
J.\,Prast,
G.\,Vouters 
\\ \it
Laboratoire d'Annecy-le-Vieux de Physique des Particules, Universit\'{e} de Savoie,
CNRS/IN2P3,
9 Chemin de Bellevue BP110, F-74941 Annecy-le-Vieux CEDEX, France
}

\author{\centering
K.\,Francis,
J.\,Repond, 
J.\,Schlereth, 
J.\,Smith$^a$, %%\endnote{Also at University of Texas, Arlington},
L.\,Xia 
\\ \it
Argonne National Laboratory,
9700 S.\ Cass Avenue,
Argonne, IL 60439-4815,
USA}

\author{\centering
E.\,Baldolemar, 
J.\,Li$^b$, %%\endnote{Deceased}, 
S.\,T.\,Park, 
M.\,Sosebee, 
A.\,P.\,White, 
J.\,Yu 
\\ \it
Department of Physics, SH108, University of Texas, Arlington, TX 76019, USA
}

\author{\centering
G.\,Eigen 
\\ \it
University of Bergen, Inst.\, of Physics, Allegaten 55, N-5007 Bergen, Norway
}

\author{\centering
Y.\,Mikami, 
N.\,K.\,Watson 
\\ \it
University of Birmingham,
School of Physics and Astronomy,
Edgbaston, Birmingham B15 2TT, UK
}
\author{\centering 
G.\,Mavromanolakis$^c$, %%\endnote{Now at CERN},
M.\,A.\,Thomson, 
D.\,R.\,Ward, 
W.\,Yan$^d$ %{Now at Dept.\ of Modern Physics, Univ.\ of Science and Technology of China, 96 Jinzhai Road, Hefei, Anhui, 230026, P.\, R.\, China}
 \\ \it
University of Cambridge, Cavendish Laboratory, J J Thomson Avenue, CB3 0HE, UK
}

\author{\centering 
D.\,Benchekroun, 
A.\,Hoummada, 
Y.\,Khoulaki
 \\ \it
Universit\'{e} Hassan II A\"{\i}n Chock, Facult\'{e} des sciences.\, B.P. 5366 Maarif, Casablanca, Morocco
}

\author{\centering 
J.\,Apostolakis, 
D.\,Dannheim, 
A.\,Dotti, 
G.\,Folger, 
V.\,Ivantchenko, 
W.\,Klempt, 
E.\,van der Kraaij$^e$, %%,now at University of Bergen 
A.\,-I.\,Lucaci-Timoce, 
A.\,Ribon, 
D.\,Schlatter, 
V.\,Uzhinskiy
 \\ \it 
CERN, 1211 Gen\`{e}ve 23, Switzerland
}

\author{\centering
C.\,C\^{a}rloganu, 
P.\,Gay, 
S.\,Manen, 
L.\,Royer
 \\ \it
Clermont Univertsit\'e, Universit\'e Blaise Pascal, CNRS/IN2P3, LPC, BP
10448, F-63000 Clermont-Ferrand, France
}

\author{\centering
M.\,Tytgat,
N.\,Zaganidis
 \\ \it
Ghent University, Department of Physics and Astronomy,
Proeftuinstraat 86, B-9000 Gent, Belgium
}

\author{\centering
G.\,C.\,Blazey,
A.\,Dyshkant, 
J.\,G.\,R.\,Lima, 
V.\,Zutshi
 \\ \it
NICADD, Northern  Illinois University,
Department of Physics,
DeKalb, IL 60115,
USA
}

\author{\centering 
J.\,-Y.\,Hostachy, 
L.\,Morin
 \\ \it
Laboratoire de Physique Subatomique et de Cosmologie - Universit\'{e} Joseph Fourier Grenoble 1 -
CNRS/IN2P3 - Institut Polytechnique de Grenoble,
53, rue des Martyrs,
38026 Grenoble CEDEX, France
}

\author{\centering 
U.\,Cornett, 
D.\,David, 
G.\,Falley, 
K.\,Gadow, 
P.\,G\"{o}ttlicher, 
C.\,G\"{u}nter,
O.\,Hartbrich, 
B.\,Hermberg, 
S.\,Karstensen, 
F.\,Krivan,
K.\,Kr\"uger, 
S.\,Lu, 
S.\,Morozov, 
V.\,Morgunov$^f$, %%\endnote{On leave from ITEP}, 
M.\,Reinecke, 
F.\,Sefkow, 
P.\,Smirnov,
M.\,Terwort
 \\ \it
DESY, Notkestrasse 85,
D-22603 Hamburg, Germany
}

\author{\centering  
N.\,Feege, 
E.\,Garutti, 
S.\,Laurien, 
I.\,Marchesini$^g$, %%\endnote{Also at DESY},
M.\,Matysek, 
M.\,Ramilli
 \\ \it
Univ. Hamburg,
Physics Depaortment,
Institut f\"ur Experimentalphysik,
Luruper Chaussee 149,
22761 Hamburg, Germany
}

\author{\centering 
K.\,Briggl, 
P.\,Eckert, 
T.\,Harion, 
H.\,-Ch.\,Schultz-Coulon, 
W.\,Shen, 
R.\,Stamen
 \\ \it
 University of Heidelberg, Fakultat fur Physik und Astronomie,
Albert Uberle Str. 3-5 2.OG Ost,
D-69120 Heidelberg, Germany
}

\author{\centering 
B.\,Bilki$^h$, %%\endnote{Also at Argonne National Laboratory},
E.\,Norbeck, 
Y.\,Onel
 \\ \it
University of Iowa, Dept. of Physics and Astronomy,
203 Van Allen Hall, Iowa City, IA 52242-1479, USA
}

\author{\centering 
G.\,W.\,Wilson
 \\ \it
University of Kansas, Department of Physics and Astronomy,
Malott Hall, 1251 Wescoe Hall Drive, Lawrence, KS 66045-7582, USA
}

\author{\centering 
K.\,Kawagoe,
Y.\,Sudo,
T.\,Yoshioka
 \\ \it
Department of Physics, Kyushu University, Fukuoka 812-8581, Japan
}

\author{\centering 
P.\,D.\,Dauncey,  
A.\,-M.\,Magnan
 \\ \it
Imperial College London, Blackett Laboratory,
Department of Physics,
Prince Consort Road,
London SW7 2AZ, UK 
}

\author{\centering 
V.\,Bartsch$^i$, %%\endnote{Now at University of Sussex, Physics and Astronomy Department, Brighton, Sussex, BN1 9QH, UK}, 
M.\,Wing
 \\ \it
Department of Physics and Astronomy, University College London,
Gower Street,
London WC1E 6BT, UK
}

\author{\centering 
F.\,Salvatore$^i$ %%\endnotemark[8]
 \\ \it
Royal Holloway University of London,
Department of Physics,
Egham, Surrey TW20 0EX, UK
}

\author{\centering 
E.\,Cortina Gil, 
S.\,Mannai
 \\ \it
Center for Cosmology, Particle Physics and Cosmology (CP3)
Universit\'{e} catholique de Louvain, Chemin du cyclotron 2,
1320 Louvain-la-Neuve, Belgium
}

\author{\centering 
 G.\,Baulieu, 
 P.\,Calabria, 
 L.\,Caponetto, 
 C.\,Combaret, 
 R.\,Della\,Negra, 
 G.\,Grenier, 
 R.\,Han, 
 J-C.\,Ianigro, 
 R.\,Kieffer, 
 I.\,Laktineh, 
 N.\,Lumb, 
 H.\,Mathez, 
 L.\,Mirabito, 
 A.\,Petrukhin, 
 A.\,Steen, 
 W.\,Tromeur, 
 M.\,Vander\,Donckt, 
 Y.\,Zoccarato 
  \\ \it
Universit\'{e} de Lyon, Universit\'{e} Lyon 1, 
CNRS/IN2P3, IPNL 4 rue E Fermi 69622,
Villeurbanne CEDEX, France
}

\author{\centering 
E.\,Calvo~Alamillo, 
M.-C.\, Fouz, 
J.\,Puerta-Pelayo 
 \\ \it
CIEMAT, Centro de Investigaciones Energeticas, Medioambientales y Tecnologicas, Madrid, Spain 
}

\author{\centering 
F.\,Corriveau
 \\ \it
Institute of Particle Physics of Canada and Department of Physics
Montr\'{e}al, Quebec,
Canada H3A 2T8
}

\author{\centering 
B.\,Bobchenko, 
M.\,Chadeeva, 
M.\,Danilov$^j$, %%\endnote{Also at MEPhI and at Moscow Institute of Physics and Technology}, 
A.\,Epifantsev, 
O.\,Markin, 
R.\,Mizuk$^j$, %%\endnotemark[9], 
E.\,Novikov, 
V.\,Popov, 
V.\,Rusinov, 
E.\,Tarkovsky
 \\ \it
Institute of Theoretical and Experimental Physics, B. Cheremushkinskaya ul. 25,
RU-117218 Moscow, Russia
}

\author{\centering 
N.\,Kirikova, 
V.\,Kozlov, 
P.\,Smirnov, 
Y.\,Soloviev
 \\ \it
P.\,N.\, Lebedev Physical Institute,
Russian Academy of Sciences,
117924 GSP-1 Moscow, B-333, Russia
}

\author{\centering 
P.\,Buzhan, A.\,Ilyin, V.\,Kantserov, V.\,Kaplin, A.\,Karakash, E.\,Popova, V.\,Tikhomirov
 \\ \it
Moscow Physical Engineering Inst., MEPhI,
Dept. of Physics,
31, Kashirskoye shosse,
115409 Moscow, Russia
}

\author{\centering 
C.\,Kiesling, 
K.\,Seidel, 
F.\,Simon$^\spadesuit$, 
C.\,Soldner, 
M.\,Szalay, 
M.\,Tesar, 
L.\,Weuste 
 \\ \it
Max Planck Inst. f\"ur Physik,
F\"ohringer Ring 6,
D-80805 Munich, Germany
}

\author{\centering 
M.\,S.\,Amjad, 
J.\,Bonis, 
S.\,Callier, 
S.\, Conforti\,di\,Lorenzo, 
P.\,Cornebise, 
Ph.\,Doublet, 
F.\,Dulucq, 
J.\,Fleury, 
T.\,Frisson, 
N.\,van der Kolk,  
H.\,Li$^k$, %%\endnote{Now at LPSC Grenoble}, 
G.\,Martin-Chassard, 
F.\,Richard, 
Ch.\,de la Taille, 
R.\,P\"oschl, 
L.\,Raux, 
J.\,Rou\"en\'e, 
N.\,Seguin-Moreau
 \\ \it
Laboratoire de l'Acc\'{e}l\'{e}rateur Lin\'{e}aire, Centre
Scientifique d'Orsay, Universit\'{e} de Paris-Sud XI, CNRS/IN2P3, BP
34, B\^atiment 200, F-91898 Orsay CEDEX, France
}

\author{\centering 
M.\,Anduze, 
V.\,Balagura, 
V.\,Boudry, 
J-C.\,Brient, 
R.\,Cornat, 
M.\,Frotin, 
F.\,Gastaldi,  
E.\,Guliyev, 
Y.\,Haddad, 
F.\,Magniette, 
G.\,Musat, 
M.\,Ruan, 
T.H.\,Tran, 
H.\,Videau
 \\ \it
 Laboratoire Leprince-Ringuet (LLR)  -- \'{E}cole Polytechnique, CNRS/IN2P3, F-91128 Palaiseau, France
}

\author{\centering 
B.\,Bulanek, 
J.\,Zacek 
 \\ \it
Charles University, Institute of Particle \& Nuclear Physics,
V Holesovickach 2,
CZ-18000 Prague 8, Czech Republic  
}

\author{\centering 
J.\,Cvach, 
P.\,Gallus, 
M.\,Havranek, 
M.\,Janata, 
J.\,Kvasnicka, 
D.\,Lednicky, 
M.\,Marcisovsky,  
I.\,Polak, 
J.\,Popule, 
L.\,Tomasek, 
M.\,Tomasek, 
P.\,Ruzicka, 
P.\,Sicho, 
J.\,Smolik, 
V.\,Vrba, 
J.\,Zalesak 
 \\ \it
Institute of Physics, Academy of Sciences of the Czech Republic, Na Slovance 2,
CZ-18221 Prague 8, Czech Republic
}

\author{\centering 
B.\,Belhorma, 
H.\,Ghazlane 
 \\ \it
Centre National de l'Energie, des Sciences et des Techniques Nucl\'{e}aires, 
B.P. 1382, R.P. 10001, Rabat, Morocco
}

\author{\centering              
K.\,Kotera, 
T.\,Takeshita, 
S.\,Uozumi
 \\ \it
Shinshu Univ.\,,
Dept. of Physics,
3-1-1 Asaki,
Matsumoto-shi, Nagano 390-861,
Japan
}

\author{\centering              
D.\,Jeans
 \\ \it
Department of Physics, Graduate School of Science, The University of
Tokyo, 7-3-1 Hongo, Bunkyo-ku, Tokyo 113-0033, Japan}

\author{{\centering 
M.\, G\"otze, 
J.\,Sauer, 
S.\,Weber, 
C.\,Zeitnitz
 \\ \it
Bergische Universit\"{a}t Wuppertal
Fachbereich 8 Physik,
Gaussstrasse 20,
D-42097 Wuppertal, Germany
}
 \\ \it
$^\spadesuit$ Corresponding author\newline
E-mail: \email{fsimon@mpp.mpg.de}
}

\author{  \\
\llap{$^a$}Also at University of Texas, Arlington\\
\llap{$^b$}Deceased\\
\llap{$^c$}Now at CERN\\
\llap{$^d$}Now at Dept.\ of Modern Physics, Univ.\ of Science and Technology of China, 96 Jinzhai Road, Hefei, Anhui, 230026, P.\, R.\, China\\
\llap{$^e$}Now at University of Bergen\\
\llap{$^f$}On leave from ITEP\\
\llap{$^g$}Also at DESY\\
\llap{$^h$}Also at Argonne National Laboratory\\
\llap{$^i$}Now at University of Sussex, Physics and Astronomy Department, Brighton, Sussex, BN1 9QH, UK\\
\llap{$^j$}Also at MEPhI and at Moscow Institute of Physics and Technology\\
\llap{$^k$}Now at LPSC Grenoble
}
\abstract{We investigate the three dimensional substructure of hadronic showers in the CALICE scintillator-steel hadronic calorimeter. The high granularity of the detector is used to find track segments of minimum ionising particles within hadronic showers, providing sensitivity to the spatial structure and the details of secondary particle production in hadronic cascades. The multiplicity, length and angular distribution of identified track segments are compared to {\sc Geant4} simulations with several different shower models. Track segments also provide the possibility for in-situ calibration of highly granular calorimeters.}

\begin{document}

\bibliographystyle{JHEP}

\section{Introduction}
\label{sec:Introduction}

The physics goals of future high-energy linear electron-positron colliders place strict requirements on the performance of the detector systems. One of these requirements is excellent jet energy resolution to provide adequate separation of gauge bosons in all-hadronic final states. Detailed simulation studies have extensively demonstrated that such resolutions can be achieved with Particle Flow Algorithms \cite{Thomson:2009rp, Brient:2002gh, pfaMorgunov}. To provide the necessary separation of particles within hadronic jets, highly granular imaging calorimeters are required. The CALICE collaboration has constructed and thoroughly tested several prototypes of electromagnetic and hadronic imaging calorimeters.

\begin{figure}
\begin{center}
  \includegraphics[width=.7\linewidth]{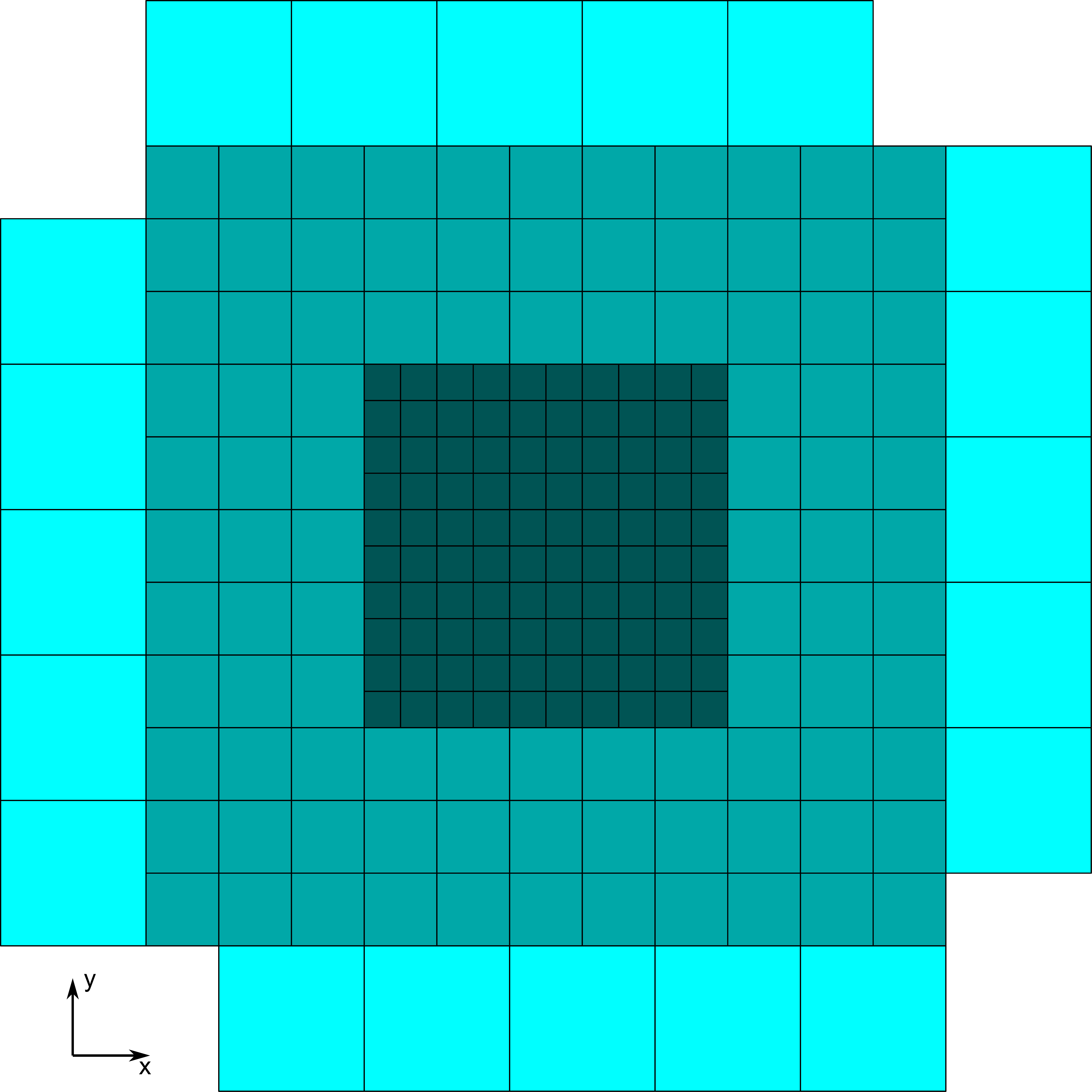}
  \caption{The cell structure of the AHCAL active layers showing the different regions of
  granularity with cells of size $\unit[30\times30]{mm^2}$,
  $\unit[60\times60]{mm^2}$ and $\unit[120\times120]{mm^2}$. }
  \label{fig:Granularity}
\end{center}
\end{figure}

One of these calorimeters is the analog hadron calorimeter (AHCAL) \cite{Adloff:2010hb}, a sampling calorimeter with 38 active layers of plastic scintillator sandwiched between steel absorber plates with lateral dimensions of  $\unit[90\times90]{cm^2}$ and a total instrumented volume of approximately $\unit[1]{m^3}$. The scintillator layers consist of \unit[5]{mm} thick scintillator tiles, each individually read out by a silicon photomultiplier (SiPM) \cite{Bondarenko:2000in, Buzhan:2003ur}. The first 30 layers of the calorimeter have a highly granular core of 10 by 10 cells each with a size of $\unit[30\times30]{mm^2}$. This is surrounded by three rings of  $\unit[60\times60]{mm^2}$ cells,
and one incomplete ring of $\unit[120\times120]{mm^2}$ cells, as shown in Figure \ref{fig:Granularity}. In the last eight layers, the highly granular core is replaced by $\unit[60\times60]{mm^2}$ cells, otherwise the layout is identical. In total, the number of cells is 7608. The passive material per layer amounts to \unit[21]{mm} of steel:   \unit[4]{mm} from the two cover plates of the cassettes housing the scintillator tiles, and \unit[17]{mm} from the absorber plates of the calorimeter structure. In total, the depth of the calorimeter is  $\unit[5.3]{\lambda_I}$, corresponding to a depth for pions of  $\unit[4.3]{\lambda_\pi}$.

During data taking at the CERN SPS, a silicon-tungsten electromagnetic calorimeter (SiW-ECAL) \cite{Anduze:2008hq} was installed upstream of the AHCAL together with a downstream tail catcher and muon tracker (TCMT) \cite{CALICE:2012aa}. These detectors are not explicitly used in the analysis presented here, but the ECAL contributes to the event selection discussed below.

\begin{figure}
\begin{center}
  \includegraphics[width=.8\linewidth]{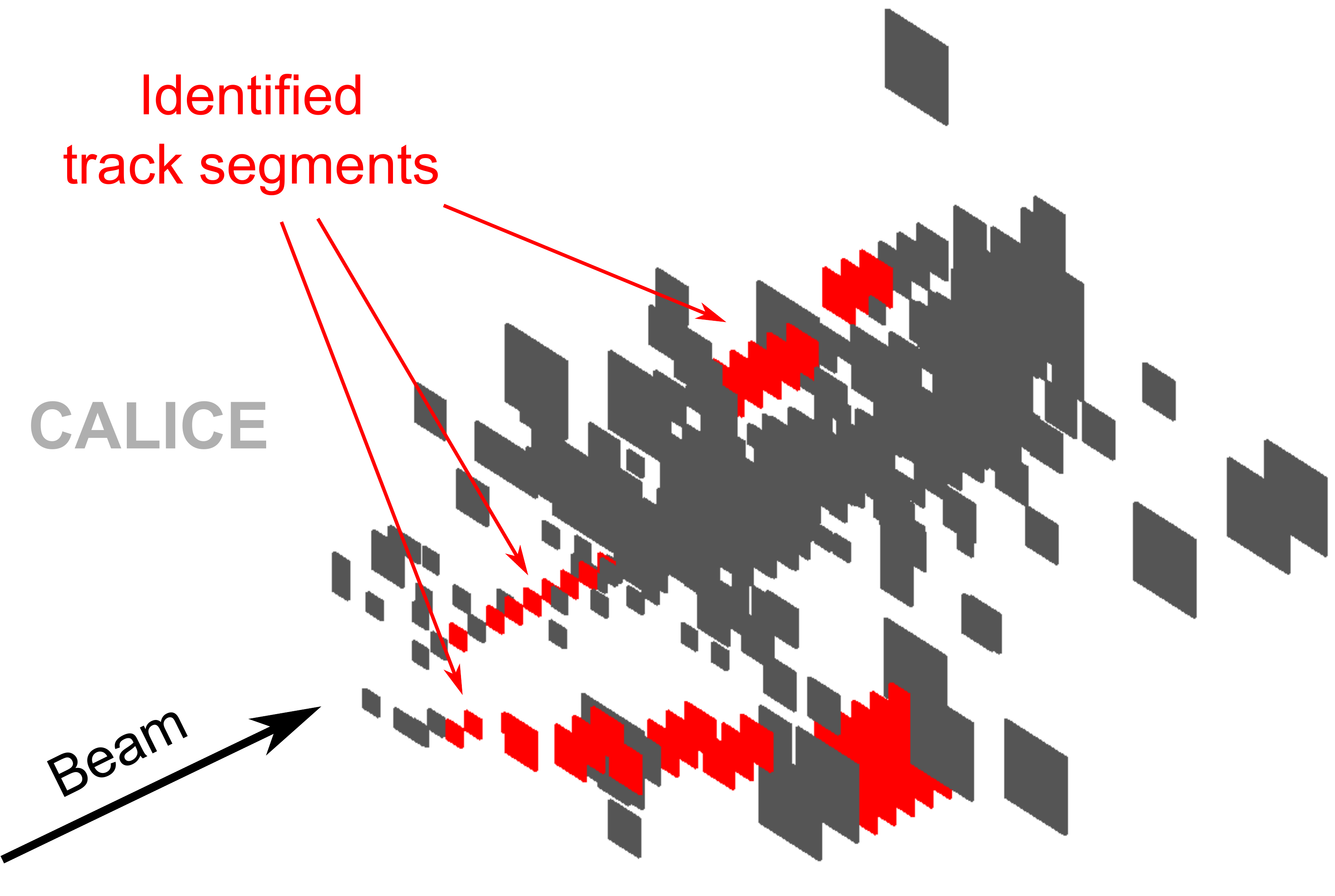}
  \caption{Event display
  of a typical hadronic shower in the  CALICE AHCAL initiated by a negative pion with an energy of 60 GeV. The identified minimum-ionising track segments are highlighted in red. The beam enters from the lower left, indicated by the black arrow.}
  \label{fig:EventDisplay}
\end{center}
\end{figure}

The high granularity of the AHCAL enables detailed investigations of the three-dimensional structure of hadronic showers. Rather than consisting of amorphous blobs of energy deposits, hadronic showers are characterised by dense electromagnetic sub-showers and sparse hadronic components forming a tree-like structure. 
The hadronic component includes (approximately) min\-i\-mum-ionising particles (MIPs) which can propagate an appreciable distance before undergoing an inelastic interaction.
In the present analysis track segments are identified using a basic nearest neighbour algorithm. 
Figure \ref{fig:EventDisplay} shows a typical hadronic shower in the AHCAL, with identified track segments highlighted in red. The properties of these track segments, such as length, multiplicity and angular distribution, are compared to {\sc Geant}4 \cite{Agostinelli:2002hh} simulations performed with several shower models.

\section{Track-finding}
\label{sec:TrackingAlgorithm}

The tracking algorithm used here consists of two stages. The first stage is the identification of track candidates in a layer by layer search using a nearest neighbour algorithm. In a second stage, these candidates are passed through a filtering algorithm based on a Hough transformation to remove  inconsistent hits such as noise hits and hits not due to energy depositions from the tracked minimum-ionising particle.

\subsection{The tracking algorithm}
\label{sec:algo:description}

\begin{figure}
\begin{center} 
  \includegraphics[width=.5\linewidth]{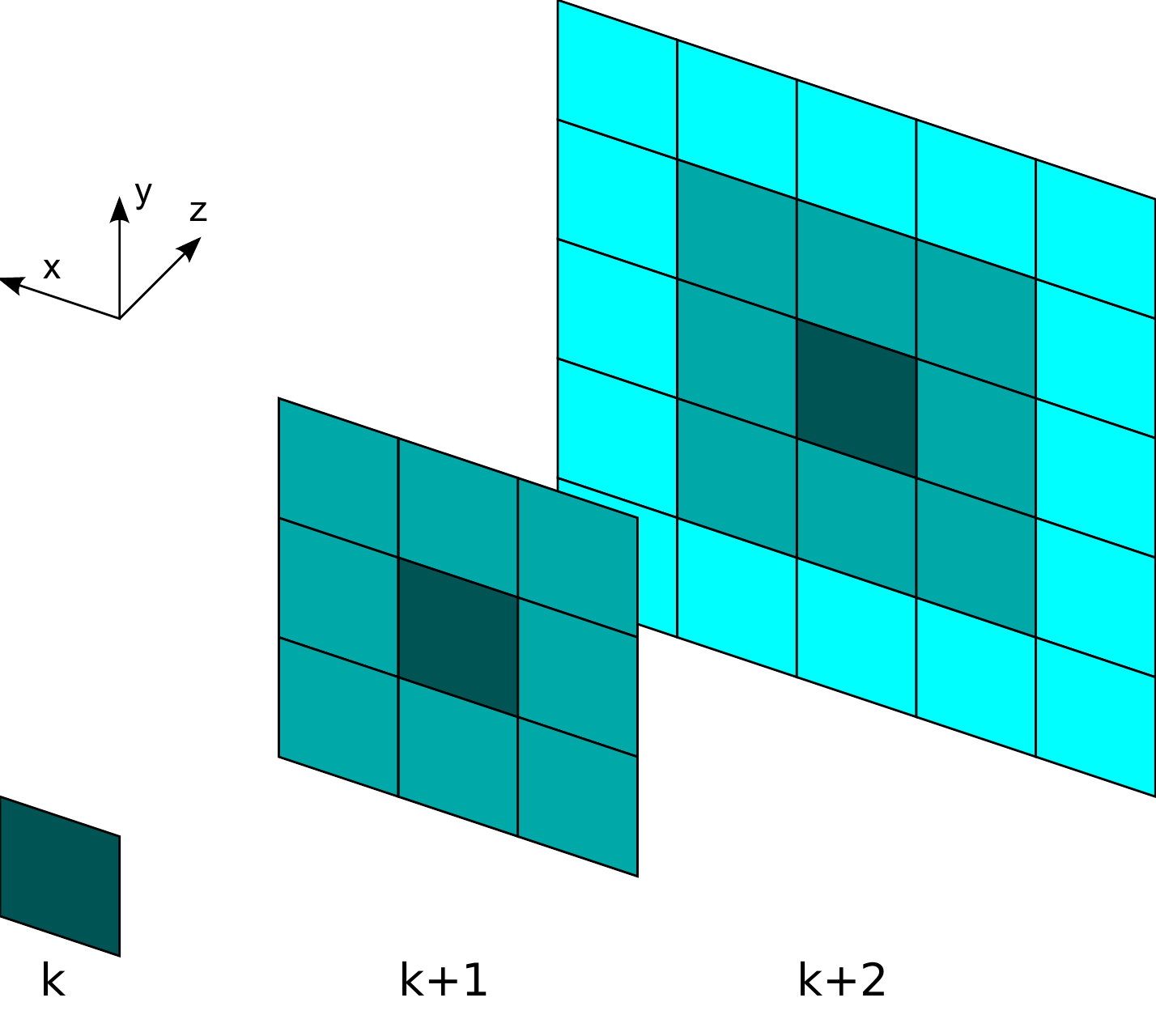}
  \caption{The search cone of the
  tracking algorithm of a cell in layer $k$. The different colours indicate the
  distance in the $x-y$ plane to the original cell.}
  \label{fig:SearchCone}
\end{center}
\end{figure}

For the track finding, the coordinate system is defined as indicated in Figure \ref{fig:SearchCone}, with the $z$-axis given by the beam axis, the $x$-axis pointing left when looking downstream in positive $z$ direction and the $y$-axis pointing up. The track finding algorithm used for the pattern recognition is a simple implementation of a nearest neighbour algorithm. The algorithm was specifically developed for the test beam data taken with the CALICE AHCAL.  It exploits the primary flight direction of incoming beam particles along the z axis by assuming that all  particles found by the algorithm have a sizeable momentum component along that axis. This is reflected by the assumption that any MIP-like particle will only create at most one hit in a given layer, and that cells on the same track in adjacent layers are neighbours, sharing at least one corner when projected on the same layer. With the  layer to layer distance of \unit[31.6]{mm} and a cell thickness of \unit[5]{mm} this limits the algorithm to the identification of tracks with a maximum angle with respect to the beam axis of approximately $60^{\circ}$ in the central region with tiles of $\unit[30\times30]{mm^2}$, of $70^\circ$ for the $\unit[60\times60]{mm^2}$ tiles and of $80^\circ$ for the outer $\unit[120\times120]{mm^2}$ tiles, respectively. It is important to note that these requirements also allow the identification of backward-going tracks, since it is not required that the tracks point outwards from the beam axis when looking downstream.

\begin{figure}[tp]
\begin{center}
  \includegraphics[width=.99\linewidth]{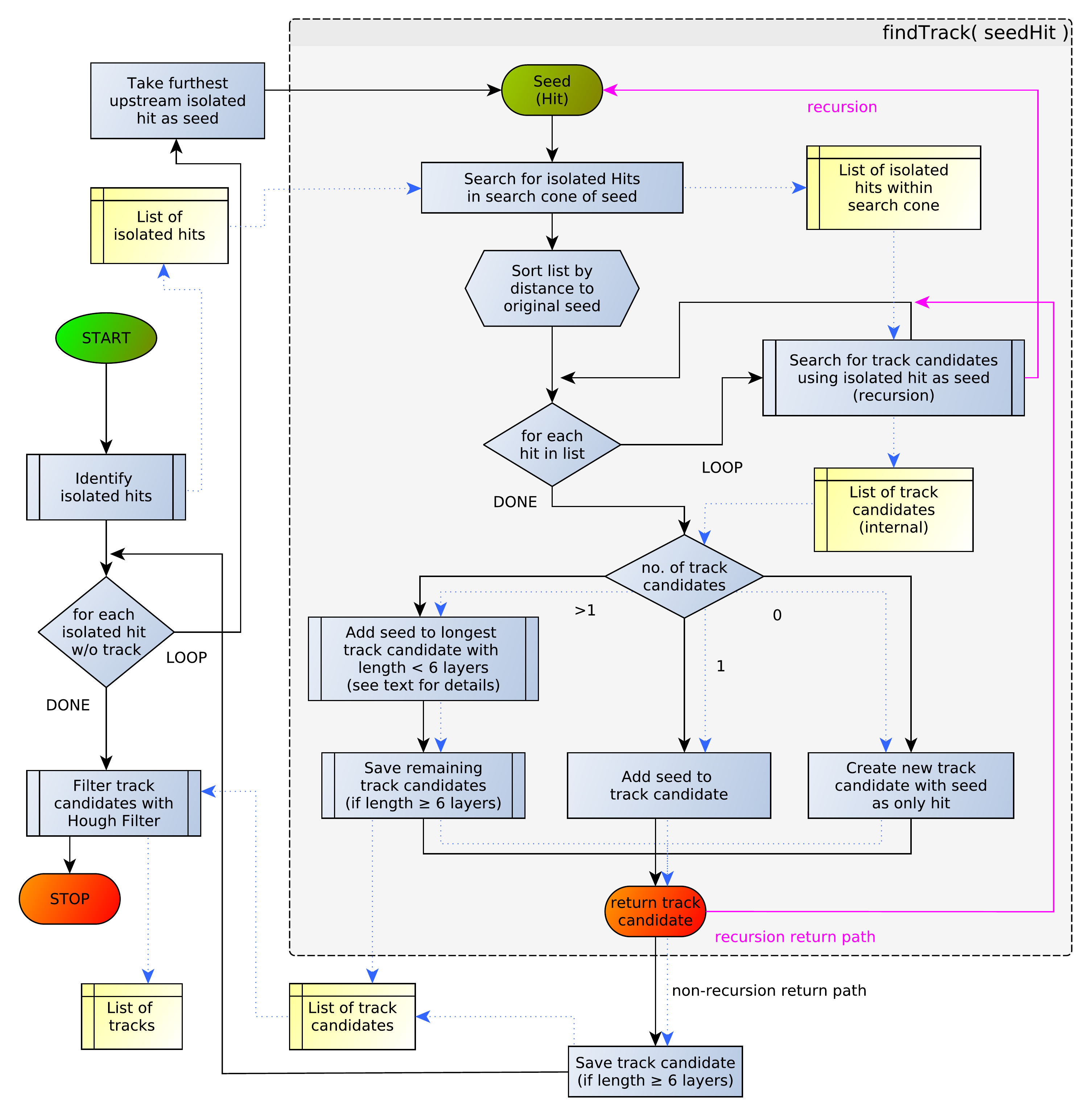}
  \caption[Tracking Algorithm Flowchart]{Flowchart of the tracking algorithm. The recursive function \emph{findTrack(seedHit)} is marked by the gray box. The control flow is indicated by the black arrows, with the exception of the recursion flow, which is indicated by the purple arrows. The blue dashed arrows show the direction of data flow for clarity.}
  \label{fig:algo:flowChart}
\end{center}
\end{figure}

Prior to the actual track finding, a set of calorimeter hits to be used in the algorithm is identified. As a first step, all hits with an amplitude below the noise cut-off are rejected. Then, isolated hits are identified. These hits have no directly adjacent - orthogonal and diagonal - neighbouring hit within the same layer. The tracking itself is performed recursively on the isolated hits only. For illustration, a flowchart of the algorithm, which is described in the following, is shown in Figure \ref{fig:algo:flowChart}, with the recursive function \emph{findTrack} enclosed in the gray box.

The underlying idea behind this recursion algorithm is that when searching for a track using a given (isolated) hit as seed, the hit itself is already a valid track candidate. A single recursion step seeks to extend this track candidate with track candidates further downstream within a search cone. These candidates in turn are identified by the (recursive) application of the same tracking algorithm, using the identified isolated hits (if they are not part of any previously identified tracks) in the search cone as seed. 
Note that the tracking algorithm only considers hits downstream with respect to the seeded hit and therefore is independent of the calling history, i.e.~the track candidates found so far.
Starting with the isolated hit which is furthest upstream, i.e.~closest to the calorimeter front,
the tracking search will be performed for all isolated hits that are not part of a previously identified track candidate.

The search cone consists of cells one layer further downstream which share at least a corner with the seeded cell when projected onto the layer of the seeded cell (i.e.~the cell with the same $x/y$ coordinates and the ring of cells around it), plus all cells two layers downstream of the seeded cell sharing at least a corner with the cells in the cone one layer downstream, again after projection onto the layer of the seeded cell (i.e.~the cell with the same $x/y$ coordinates and two rings of cells around it). This search cone is illustrated in Figure \ref{fig:SearchCone}. Extending the search cone over two layers has the advantage that it gives the tracking algorithm the possibility to bridge gaps in a track, which can originate from the finite efficiency of the scintillator cells and from individual cells failing the isolation criterion due to neighbouring noise hits or other energy deposits.

All isolated hits within this cone which are not part of any track candidate so far are identified. As hits can only be assigned to one track candidate, all these hits are sorted by distance to the seeded cell to favor track continuity over gaps. As stated before, for each of these hits another recursion of the tracking algorithm is started and the resulting track candidates, each of which is a viable option to extend the current track candidate consisting of only the seeded cell, are collected. 
At this stage, three possibilities are considered:
\begin{itemize}
\item There is no isolated hit within the search cone of the seeded cell. Hence the seeded cell is the last one (furthest downstream) on a possible track. Therefore the recursion returns a new track candidate with the seeded cell as its only member.
\item If a single isolated hit, i.e.~a single track candidate is found within the search cone of the seeded cell, the seeded cell itself is added to the candidate and the track candidate is returned, ending this recursion step.
\item If the recursive calls return more than one track candidate as a possible continuation from the seeded cell, it has to be decided to which of these candidates the seeded cell should be added to. For this analysis it was decided to maximize the number of found tracks. 
Therefore all track candidates that already surpass the minimum length threshold for the acceptance of tracks of five layers are declared final and added to the list of identified tracks, 
while the seeded cell is added to the longest candidate below the threshold. However, if all track candidates surpass the threshold, the shortest one is chosen to accept the additional hit (and thus not declared final yet). 
This case of multiple possible continuations from the seeded cell is  relatively rare, and occurs in less than 9\% of the cases where a continuation is found.
\end{itemize}
At the completion of the algorithm, once all recursions are closed, the remaining uncompleted track candidates are declared to be final tracks if they fulfil the minimum length requirement of five layers.

\subsection{Hough based track filtering}
\label{sec:algo:filter}

The tracking algorithm presented in Section \ref{sec:algo:description} works on a simple nearest neighbour basis and does not take continuity of the track direction into account. Noise hits and other energy deposits not originating from the tracked particle which are added to a track can give it an unphysical kink or step. In rare cases a track can even entirely consist of noise and other random hits. To remove hits probably not belonging to tracks and to eliminate fake tracks, a filtering stage is applied after track finding. Since fit based solutions with $\chi^2$-based straight line fits were found to be inefficient in this analysis, the used filter is based on a two-dimensional Hough transformation \cite{Fruehwirt:DataAnalysis:Hough}. It identifies hits which are inconsistent with the dominant direction of the track. In the filter, the $x-z$ plane and the $y-z$ plane are treated separately. In the following, the procedure for the $x-z$ plane is described in detail. The $y-z$ plane is treated in an analogous way.

As the maximum track angle is limited to less than 90$^\circ$ with respect to the beam axis, a track given by a straight line can be described by an equation of the form given in Equation  \ref{eq:lineNorm}. Each point in real space transforms into Hough space (described by the coordinates $t$ and $m$) into a line as given by Equation \ref{eq:lineHough}.
\begin{eqnarray}
x & = & mz + t \label{eq:lineNorm} \\
t & = & -zm + x \label{eq:lineHough}
\end{eqnarray}
Points lying on one common track transform into lines which intersect in Hough space.  The coordinates of this intersection are the parameters $m$ and $t$ of the original Equation \ref{eq:lineNorm} for an identified track.

Since the cells of the calorimeter have a finite size, the $x$ and $y$ position of a hit is not known precisely, but is uncertain by the horizontal and vertical size of the cell of \unit[30]{mm}, \unit[60]{mm} or \unit[120]{mm}. Thus, a hit $(x,z)$ in a cell with size $s$ is transformed to a band in the Hough space, with the upper and lower limits given by 
\begin{eqnarray}
t & = & -mz + x  \pm s/2 \label{eq:lineHoughBandUpper}.
\end{eqnarray}
The problem of finding hits on a straight line through the intersections of lines in Hough space thus transforms into one of finding areas of overlap of the individual bands. The regions of overlap are determined numerically. The area shared by the highest number of bands is taken as the range of track parameters in Hough space. All hits whose bands are not overlapping with this area are considered to be noise and are removed from the track. If the resulting cleaned track has fewer than four hits 
the entire track is removed from the event.

\section{Data set and detector simulations}
\label{sec:EventSelection}

The tracks identified in hadronic showers are used to validate the realism of detector simulations based on {\sc Geant4} by comparing results from test-beam data and simulated data. The analysis is based on a data set recorded at the CERN SPS at the H6 beam line in 2007 with energies ranging from \unit[10]{GeV} to \unit[80]{GeV}. All runs included in the present analysis used negative polarity secondary or tertiary beams, consisting of a mix of negative pions, muons and electrons. Antiproton contamination as well as kaon content are negligible. 

\subsection{Calibration}

The response of each calorimeter cell is calibrated with high-energy muons, using the visible signal of these minimum-ionising particles (MIP) as the cell-to-cell calibration scale \cite{Adloff:2010hb}. Throughout the event selection and the analysis, this calibration scale is used since no conversion to reconstructed hadronic energy is required. 

\subsection{Event selection}

\begin{figure}[tp]
\begin{center}
  \includegraphics[width=.80\linewidth]{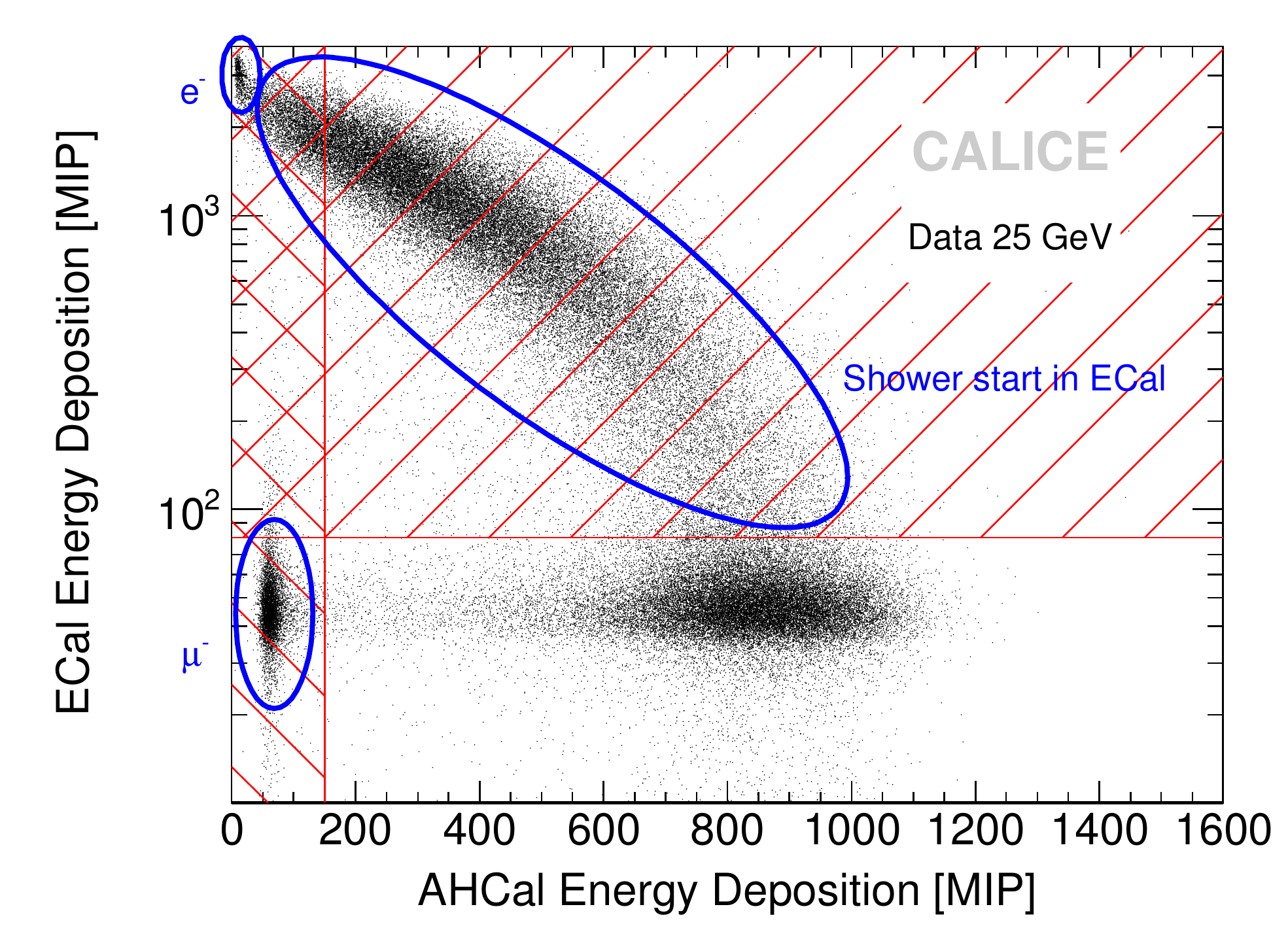}
  \caption[Event Selection Cut]{The energy of the ECal versus the energy of the AHCal of a \unit[25]{GeV} hadron run. Regions with electron and muon content as well as events with showers starting in the ECAL are marked in blue, while the event selection cuts are indicated in red.}
  \label{fig:EventSelection:MipCut}
\end{center}
\end{figure}

With the event selection procedure, a pion sample of high purity entering the AHCAL is selected. Figure \ref{fig:EventSelection:MipCut} shows all events of a \unit[25]{GeV} run with mixed particle content as a scatter plot of the energy measurement of the hadronic and the electromagnetic calorimeters. Within this plot different particle types are marked.
Electrons are eliminated by rejecting events with an energy deposition of more than \unit[80]{MIP} in the ECAL upstream of the AHCAL, which is incompatible with the assumption of a minimum-ionising particle traversing the entire ECAL. This requirement also rejects most events in which the pion has its first inelastic interaction in the ECAL. In addition, a minimum energy deposition of \unit[150]{MIP} in the AHCAL is required, which is approximately three times the
energy deposited by a minimum-ionising particle traversing the entire AHCAL. This selection rejects muons and requires some hadronic shower activity to take place in the calorimeter. To illustrate these cuts, Figure \ref{fig:EventSelection:MipCut} shows the distribution of reconstructed energy in the ECAL versus reconstructed energy in the AHCAL for a \unit[25]{GeV} hadron run, with the regions rejected by the selection cuts shown in red. The  cuts successfully select hadrons showering in the AHCAL, while rejecting muons, electrons (which are fully absorbed in the ECAL) and showers starting already in the ECAL. Prior to the event selection this data set contains approximately $\unit[1.2]{\%}$ electrons and $\unit[5.8]{\%}$ muons, which are both successfully suppressed to a negligible contribution after the event selection. Since approximately 60\% of all hadrons already start showering in the ECAL, about  $\unit[35]{\%}$ of all events are accepted in the analysis, corresponding to typically 50\,000 to 80\,000 events at each energy.

In each event, noise hits are rejected by requiring a minimum amplitude of  \unit[0.5]{MIP} per channel. After this cut, the remaining noise contribution, measured in random trigger events, is $\unit[8.88 \pm 0.14]{MIP}$ per event.

\subsection{Detector simulations}

The simulations are based on a detailed model of the test-beam setup including calorimeters and beamline instrumentation performed with {\sc Geant}4~9.4p02. Following the {\sc Geant4} shower simulations, detector-specific effects such as the light-yield of the scintillator, the response of the SiPMs and the influence of the readout electronics are simulated in a digitisation stage. Detector noise is taken into account by overlaying noise recorded at the test beam with random triggers during the runs that are to be compared with the simulations. The dependence on environmental parameters is fully taken into account. Both data and simulations are treated identically by the reconstruction code which identifies hits using the MIP-scale calibration. Further details on the simulation procedure of the AHCAL can be found in \cite{collaboration:2010rq}. At each energy, 150\,000 events are simulated, resulting in typically 65\,000 accepted events.

{\sc Geant}4 provides various models for the simulation of hadronic interactions with different regions of validity in energy and for different particle species. To cover the full energy range of interest for detector simulations in high-energy physics, so-called physics lists are provided which are combinations of several of these models \cite{Geant4:PhysicsLists, Geant4:PhysicsLists:Improvements}. In energy regions where two models are overlapping in the physics list, {\sc Geant}4 chooses
one of the two randomly on a call-by-call basis. The probability for the model valid at lower (higher) energies to be chosen decreases (increases) linearly with energy in the overlap region to provide a smooth transition between models.

For the analysis in this paper, the following four physics lists are used:

\begin{description}
  \item[LHEP]  Uses the LEP (low energy parametrised) and HEP (high energy parametrised) models, with a transition region from \unit[25]{GeV} to \unit[50]{GeV}. This physics list is essentially a {\sc Geant}4 adaptation of the GEISHA model \cite{Fesefeldt:1985yw} used in {\sc Geant}3. It is known to be less accurate than newer models, but is included here to provide an indication of the progress achieved with more recent codes.

  \item[QGSP\_BERT] Uses a Quark Gluon String (QGS) model \cite {Folger:2003sb} followed by the Precompound (P) and evaporation model for the de-excitation of nuclei for energies above \unit[12]{GeV}. The Bertini (BERT) cascade \cite{Heikkinen:2003sc} is used for energies below \unit[9.9]{GeV}. In the intermediate region between those two models in the range from \unit[9.5]{GeV} to \unit[25]{GeV} the LEP model is used. 

  \item[FTFP\_BERT] Uses the Fritiof (FTF) model followed by a Reggeon cascade
  and the Precompound evaporation (P) model \cite{NilssonAlmqvist:1986rx} for energies higher than
  \unit[4]{GeV}. Below \unit[5]{GeV} the Bertini cascade is used. This physics list uses the same cross section model
  as the QGSP\_BERT list. 

  \item[QGS\_BIC] This list is identical to QGSP\_BERT for energies above
  \unit[12]{GeV}.  However, for lower energies the Bertini cascade is
  replaced by a combination of the LEP model and the binary cascade (BIC) \cite{BIC}, with a transition between \unit[1.2]{GeV} and \unit[1.3]{GeV}.

\end{description}

\section{Results}

Since the tracking algorithm introduced in Section \ref{sec:algo:description} uses only isolated hits, it finds MIP-like track segments which are well separated from regions of dense shower activities. Apart from the track of the incoming charged pion prior to its first inelastic interaction, these are mainly tracks of higher-energy secondary particles which travel an appreciable distance before interacting again. A comparison of the test beam results to simulations is thus most sensitive to the sparse outer and tail regions of hadronic showers. To disentangle possible differences between the primary track of the incoming hadron and secondary tracks, the energy dependence of the observables are studied both for all identified tracks and tracks starting in layer three or later.

\subsection{Track multiplicity}

\begin{figure}
\begin{center}
  \includegraphics[width=.7\linewidth]{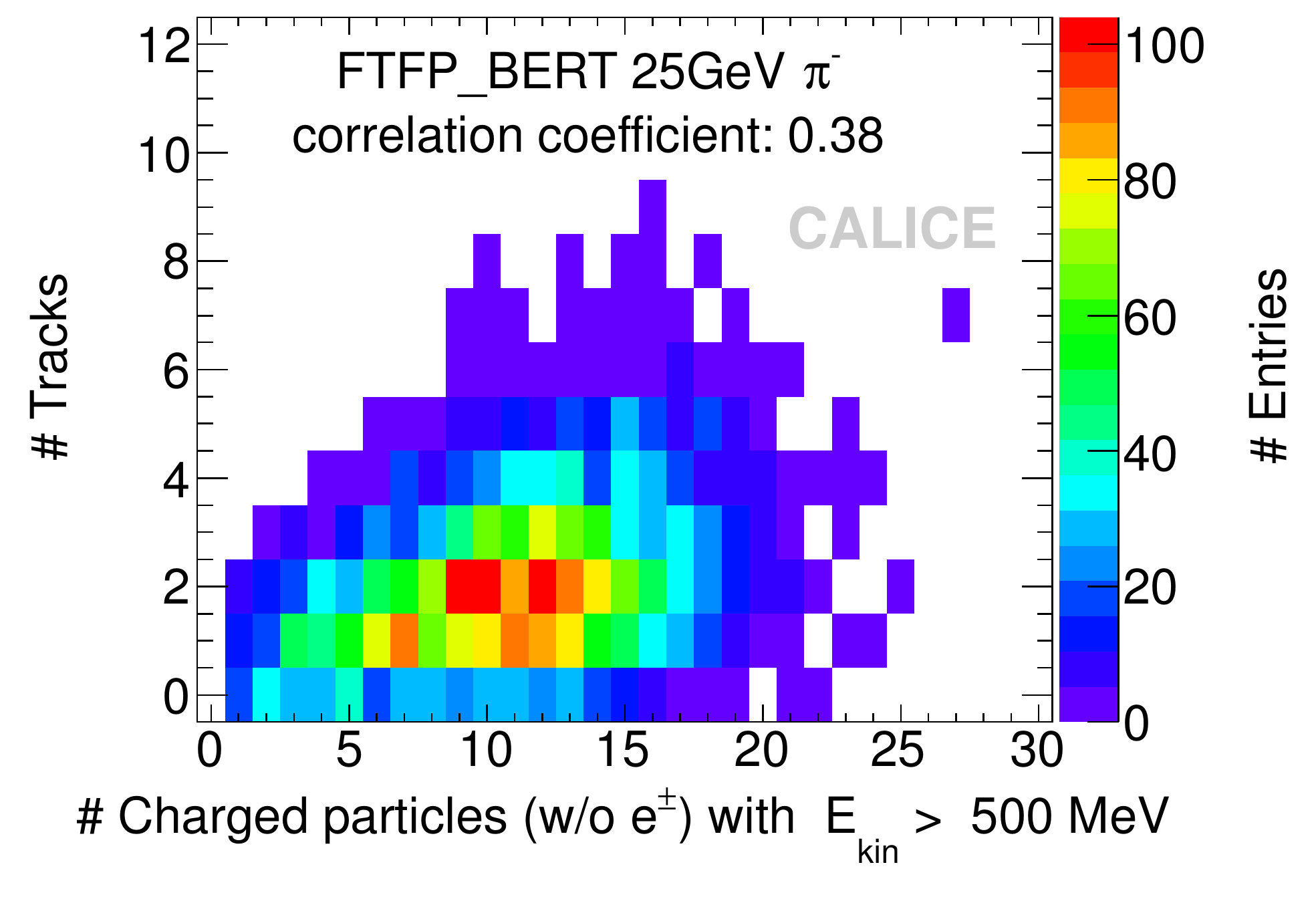}
  \caption{Correlation between the number of identified track segments and the true number of charged particles (excluding electrons and positrons) with a kinetic energy of greater than \unit[500]{MeV}. These results were obtained with the FTFP\_BERT physics list for 10000 simulated events at an energy of \unit[25]{GeV}. The Pearson correlation coefficient of the distribution is $0.38$.
  }
  \label{fig:multiplicityCorrelation}
\end{center}
\end{figure}

The multiplicity of identified tracks per event is sensitive to the number of secondary isolated charged hadrons created in the hadronic shower.  Figure \ref{fig:multiplicityCorrelation} shows the correlation between the number of identified track segments and the true number of charged particles (except electrons and positrons) with a kinetic energy greater than \unit[500]{MeV} for events simulated with the FTFP\_BERT physics list for a beam energy of \unit[25]{GeV}. The threshold of \unit[500]{MeV} is chosen to ensure that pions remain approximately minimum-ionizing for the typical length of identified tracks despite ionization energy loss in the absorber plates. For both data and simulations the number includes the track of the primary pion. The observed correlation demonstrates that the multiplicity of identified track segments is indeed sensitive to the overall number of energetic secondary particles.  The Pearson correlation coefficient has a value of 0.38 for FTFP\_BERT and of 0.41 for QGSP\_BERT. 

\begin{figure}
\begin{center}
  \includegraphics[width=.80\linewidth]{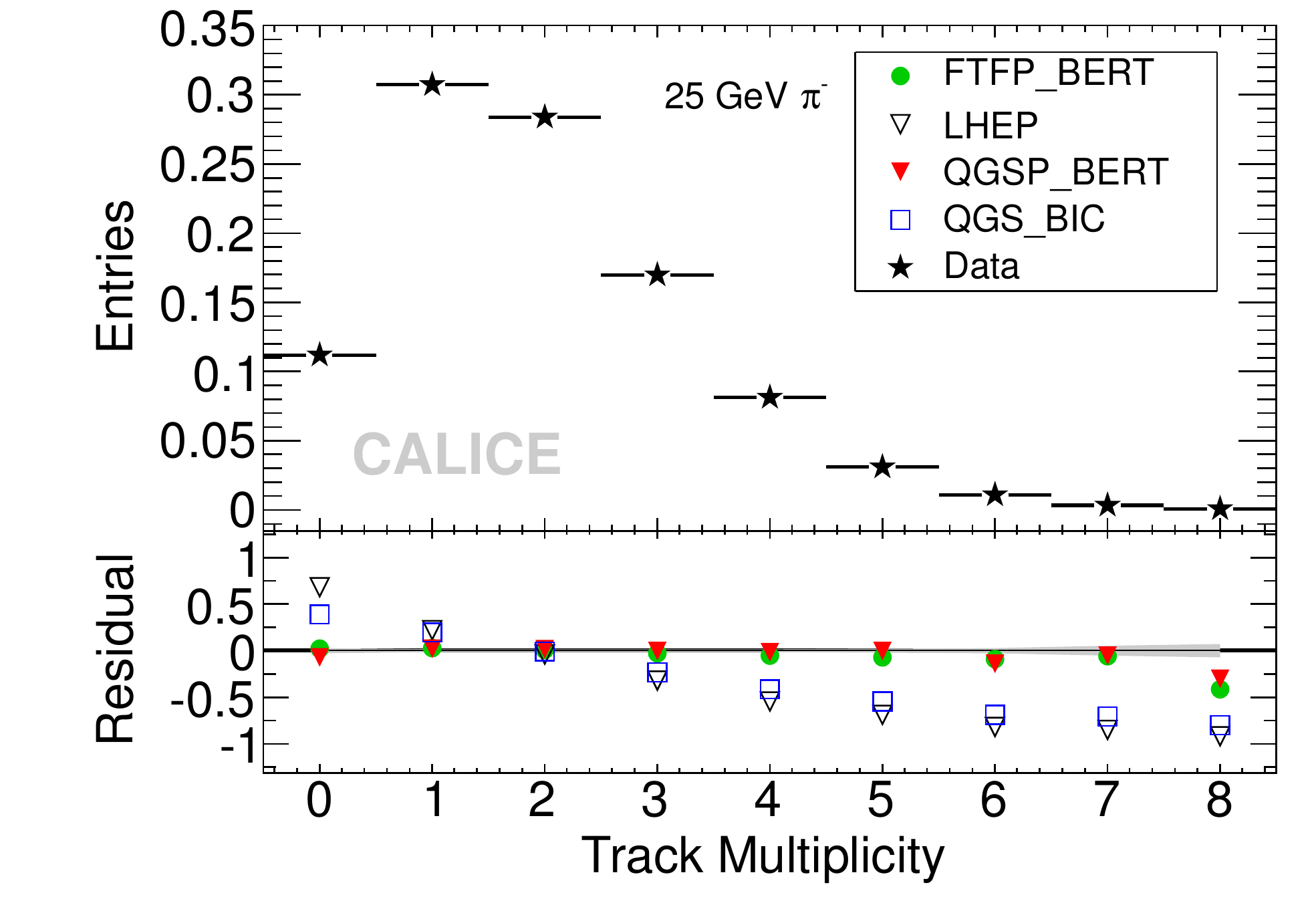}
    \caption{
    Distribution of track multiplicity for \unit[25]{GeV} pion showers. 
    The upper panel shows the normalised distribution for test beam data, while the lower panel shows the normalised residuals (${\rm simulation}/{\rm data}-1$) between test beam data and the different physics lists. 
    The grey area indicates the statistical error of the residual between test beam data and QGS\_BIC.}
  \label{fig:MCData:TrackMult:25GeV}
\end{center}
\end{figure}

The multiplicity of identified tracks for data taken with an energy of \unit[25]{GeV} is shown in the upper panel of Figure \ref{fig:MCData:TrackMult:25GeV}. The comparison to simulations with different physics lists is shown in the lower panel as the residual $r$ between data and simulations, given by $r = \frac{\textrm{simulations -
data}}{\textrm{data}}$. Here, the residual is calculated for distributions normalised to the overall integral to compare the shape of the distributions rather than the overall number of tracks.  The grey band indicates the statistical uncertainty of the residual of
data and simulations with the QGS\_BIC physics list to give an indication of the size of the statistical uncertainties in the study. Since the number of simulated events is identical for all physics lists, the statistical errors of the other data points are omitted here and for the rest of this article for better legibility. In comparison to data, the four physics lists studied fall into two groups. While the QGSP\_BERT and the FTFP\_BERT lists reproduce the distribution of the track multiplicity very well, simulations with the LHEP and the QGS\_BIC physics lists  show significant discrepancies, with an overall shift towards a lower number of identified tracks. 

\begin{figure}
\begin{center}
  \includegraphics[width=.80\linewidth]{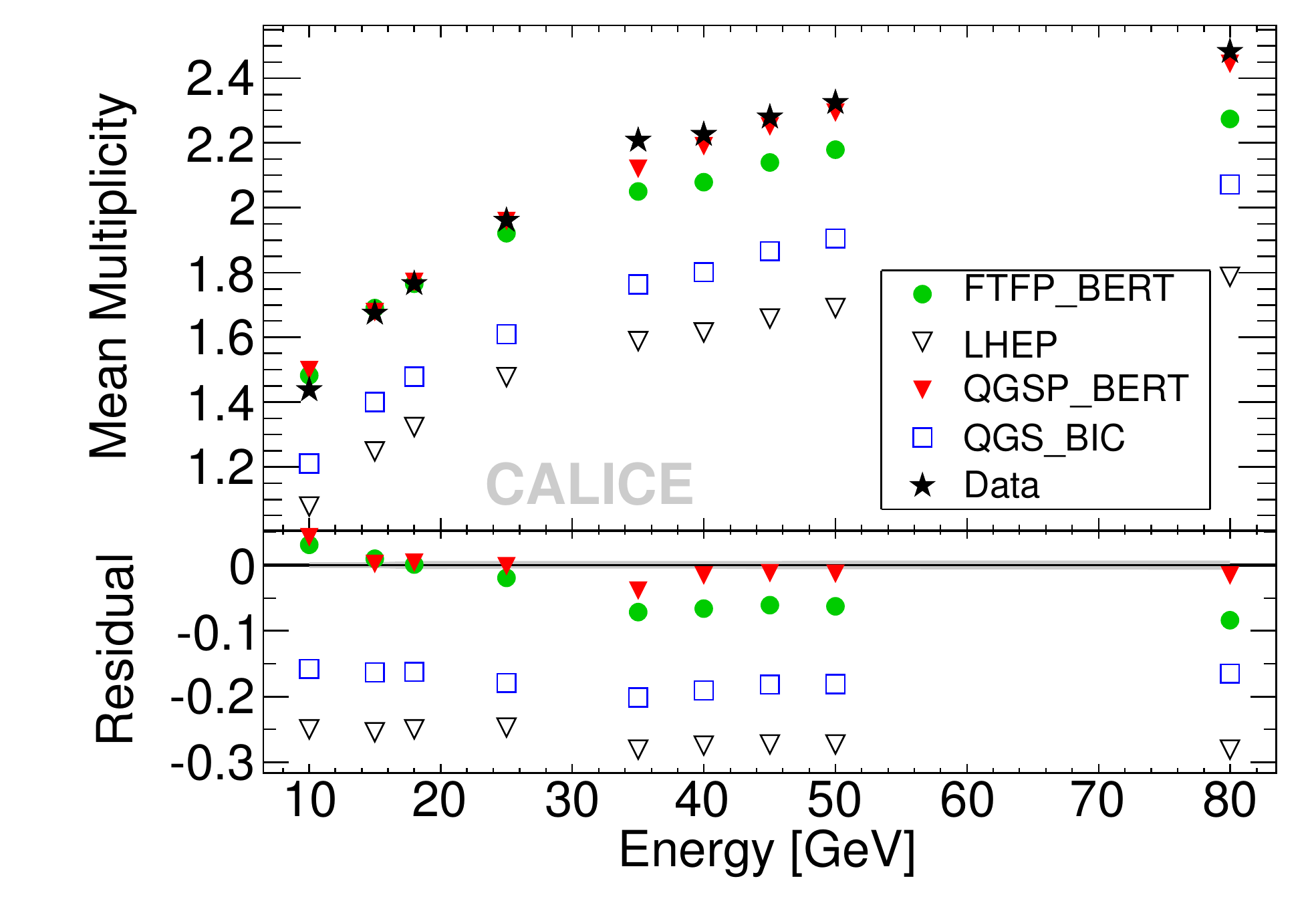}
  \caption{Mean track multiplicity as a function of energy. The upper panel shows data while the lower one shows the normalised residuals  (${\rm simulation}/{\rm data}-1$)
  between test beam data and the different physics lists. The grey area
  indicates the statistical error of the residual of test beam data and
  QGS\_BIC. Systematic errors are below the level of statistical errors, as discussed in Section \protect\ref{sec:Systematics}, and are not shown.}
  \label{fig:MCData:TrackMult:mean}
\end{center}
\end{figure}

This comparison is further expanded by a study of the energy dependence of the mean track multiplicity per event, shown in Figure \ref{fig:MCData:TrackMult:mean}. Again, the statistical errors are omitted except for the residual of data and simulations with QGS\_BIC. As for the differential distribution discussed above, the QGSP\_BERT and the FTFP\_BERT lists reproduce the energy dependence of the mean multiplicity quite well, with QGSP\_BERT showing the best agreement with deviations below 5\% at all energies. Simulations with QGS\_BIC consistently underestimate the number of tracks by 15\%, and LHEP produces 25\% to 30\% too few tracks. This shows that these two lists, in particular the rather old LHEP model, provide insufficient production of higher energy secondary hadrons which propagate outside of the core of the shower. Since QGS\_BIC uses  the low-energy part of the LHEP model, LEP, for mesons in a wide energy range, its poorer performance compared to QGSP\_BERT and FTFP\_BERT is expected in light of the observed LHEP performance.

\begin{figure}
\begin{center}
  \includegraphics[width=.80\linewidth]{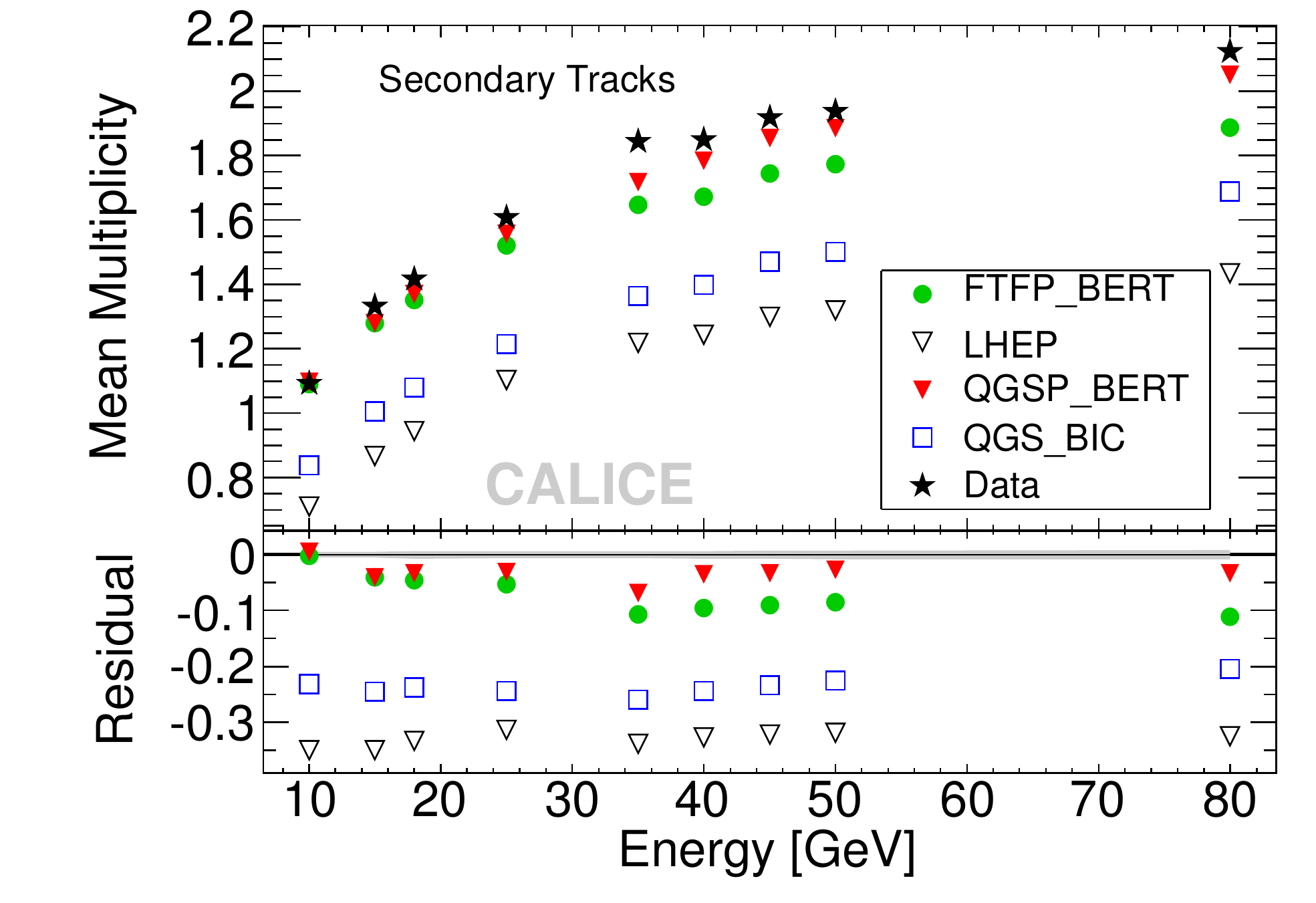}
  \caption{Mean secondary track multiplicity (tracks starting in layer three or later) as a function of energy. The upper panel shows data while the lower one shows the normalised residuals  (${\rm simulation}/{\rm data}-1$)
  between test beam data and the different physics lists. The grey area
  indicates the statistical error of the residual of test beam data and
  QGS\_BIC. Systematic errors are below the level of statistical errors, as discussed in Section \protect\ref{sec:Systematics}, and are not shown.}
  \label{fig:MCData:TrackMult:meanSec}
\end{center}
\end{figure}

Consistent behavior is observed when considering only secondary tracks, defined here as starting in calorimeter layer three or later. Figure \ref{fig:MCData:TrackMult:meanSec} shows the mean multiplicity of identified secondary tracks as a function of energy. Compared to the full track sample shown in Figure \ref{fig:MCData:TrackMult:mean} the mean is reduced, as expected from the exclusion of primary track segments. The comparison to simulations show a slight increase of the discrepancy between data and the QGS\_BIC and LHEP models with respect to the observations for the inclusive track sample, consistent with the interpretation of insufficient production of higher energy secondary hadrons.

\subsection{Track inclination}

\begin{figure}
\begin{center}
  \includegraphics[width=.80\linewidth]{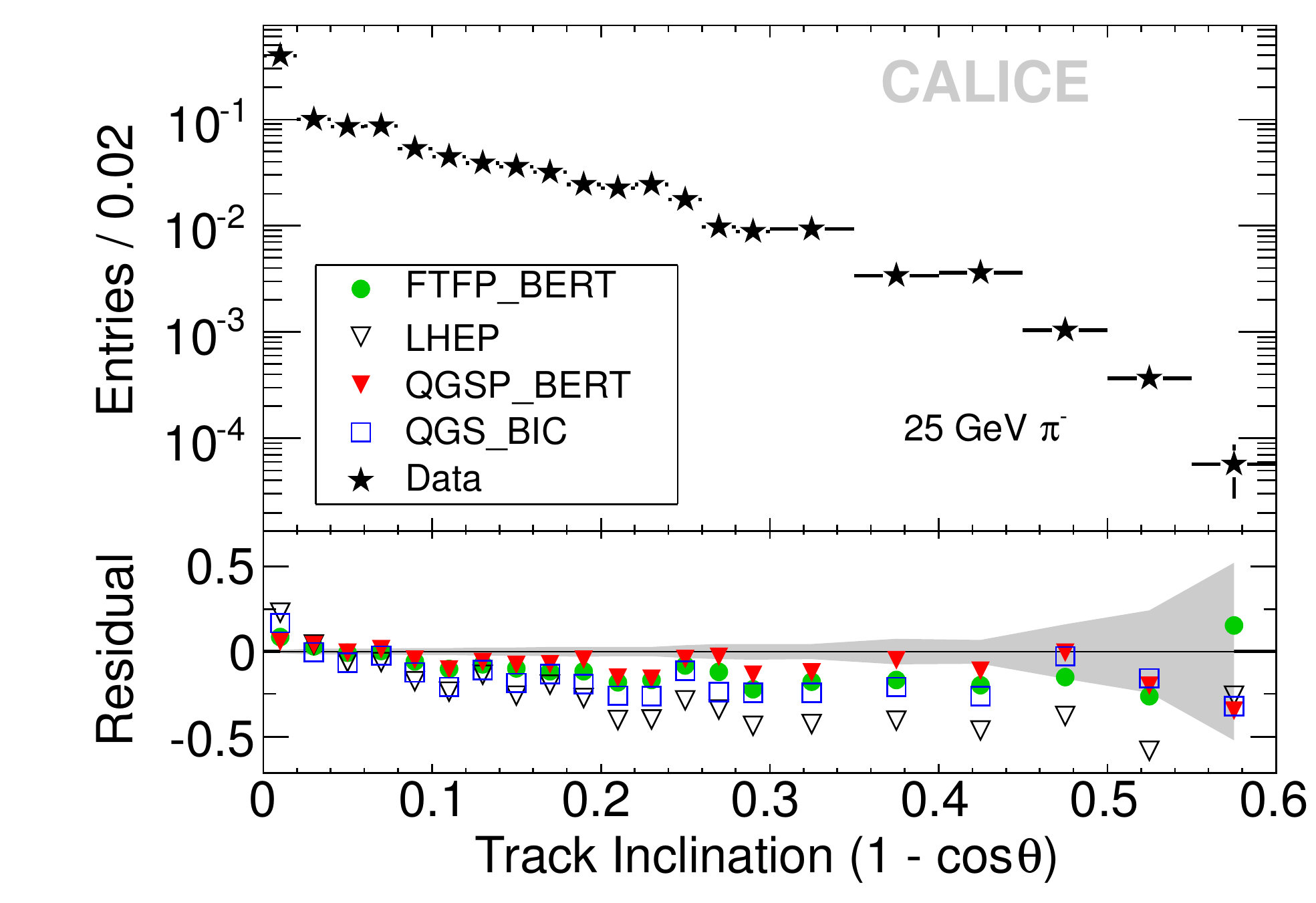}
  \caption{Normalised distribution of inclination of identified tracks for \unit[25]{GeV} pion showers. The upper panel shows the distribution for test beam
  data normalised to an integral of 1, while the lower panel shows the normalised residuals  (${\rm simulation}/{\rm data}-1$)
  of data and the different physics lists. The grey area indicates the
  statistical uncertainty of the residual of data and the QGS\_BIC physics list. Systematic errors are below the level of statistical errors, as discussed in Section \protect\ref{sec:Systematics}, and are not shown.}
  \label{fig:MCData:TrackCosPhi:25GeV}
\end{center}
\end{figure}

The distribution of $1 - \cos\theta$, where $\theta$ is the angle of the track with respect to the beam axis, here for brevity referred to as track inclination, provides sensitivity to the contribution of higher-energetic secondary particles emitted at large angles. The inclination of a track is calculated by dividing the distance in $z$ between the position of the centre of the first and the last
tile in the track by their absolute distance $r$,
\begin{eqnarray}
1 - \cos \theta & = & 
1 - \frac{\Delta z}{\Delta r} = 1 -
\frac{\Delta z}{\sqrt{\Delta x^2 + \Delta y^2 + \Delta z^2}}.
\end{eqnarray}
The algorithm is capable of identifying tracks with an inclination of 0 up to  
$1 - \cos \theta = 0.47$ for the central detector region and up to $1 - \cos \theta = 0.69$ for the
outer detector region. 

Figure \ref{fig:MCData:TrackCosPhi:25GeV} shows the distribution of the track inclination for \unit[25]{GeV} pion showers, with the lower panel showing the comparison to simulations. The integral of the distributions is normalised to unity to provide sensitivity to the shape of the distribution while ignoring the influence of differences in the overall number of identified tracks between data and simulations with different physics lists. The number of identified tracks falls approximately exponentially with higher inclination. An exception to this is the bin from $0$ to $0.02$ which contains an excess of tracks since it also includes the tracks of the primary beam particles.  While all physics lists follow this trend, most tend to underestimate the importance of tracks with high inclinations emitted at large angles. Again the QGSP\_BERT physics list provides a good description of the data while the relative importance of large-angle tracks is underestimated by approximately a factor of two by the LHEP physics list.

 \begin{figure}
\begin{center}
  \includegraphics[width=.80\linewidth]{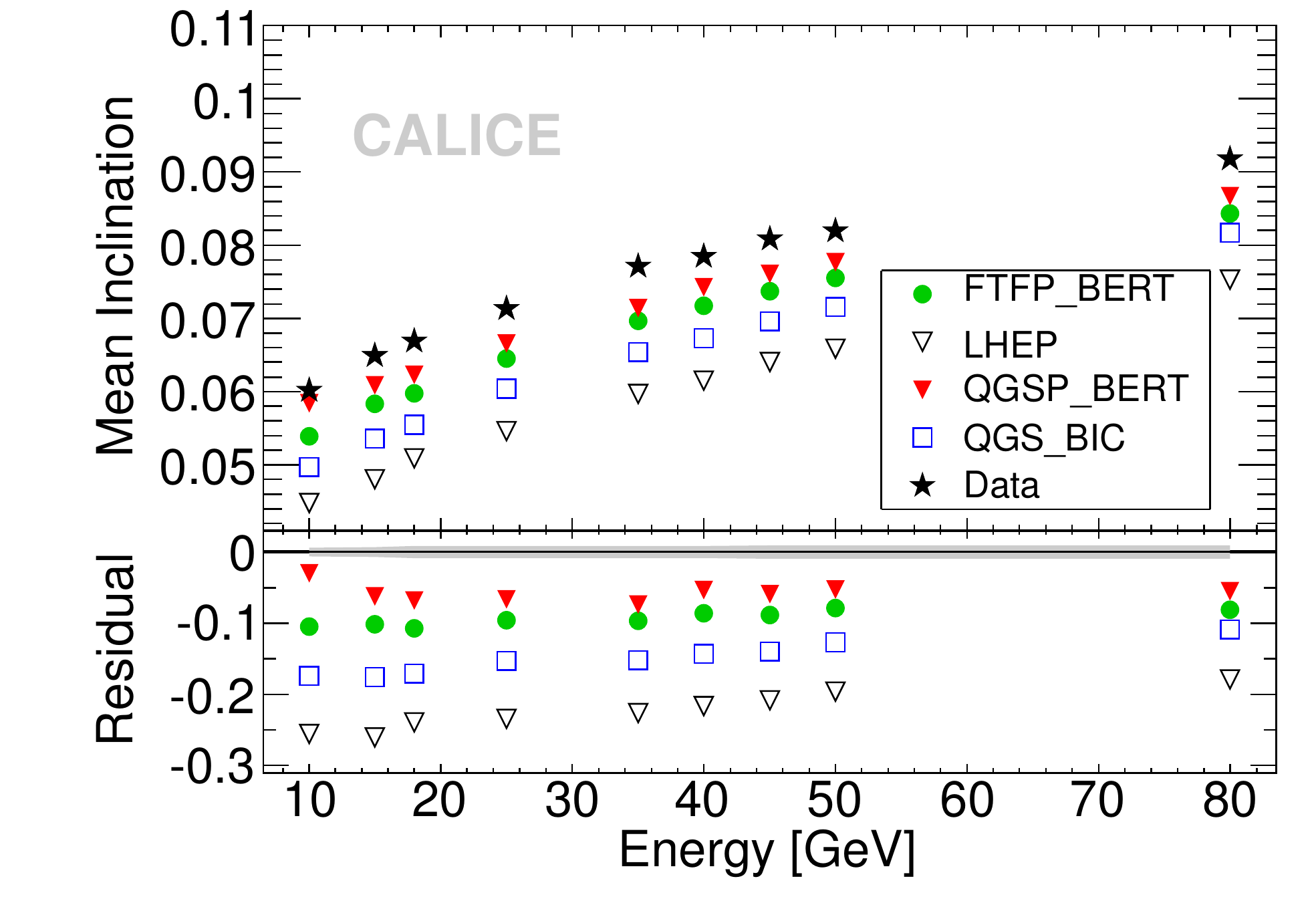}
  \caption{Mean track inclination as a function of energy. The upper panel shows the
  mean track inclination ($1-\cos \theta$) for all physics lists
  and test beam data. The lower panel shows the normalised residuals  (${\rm simulation}/{\rm data}-1$) of data and simulations with the grey area showing the statistical uncertainty of the residual of data and QGS\_BIC. Systematic errors are below the level of statistical errors, as discussed in Section \protect\ref{sec:Systematics}, and are not shown.}
  \label{fig:MCData:TrackCosPhi:Mean}
\end{center}
\end{figure}

This is further illustrated by the energy dependence of the mean track inclination shown in Figure \ref{fig:MCData:TrackCosPhi:Mean}. The average track inclination increases with increasing energy, meaning that the contribution of large-angle tracks to the overall track sample increases. This is consistent with the observed increase in overall track multiplicity as discussed above, which is connected to an increase in the number of identified secondary tracks which are not necessarily aligned with the beam direction. This trend is reproduced by all physics lists. As expected from the observed differences in the shape of the track angle distribution, the quality of the description of the mean track inclination observed for data varies from list to list. In general, the differences are small due to the strong predominance of tracks at inclination values close to zero. Simulations with the QGSP\_BERT physics list, which gives the best description, agree with the data at a level better than 5\%, and in particular at low energies agree well with observations. On the other hand, LHEP, which gives the worst description, shows deviations of up to 20\%.

 \begin{figure}
\begin{center}
  \includegraphics[width=.80\linewidth]{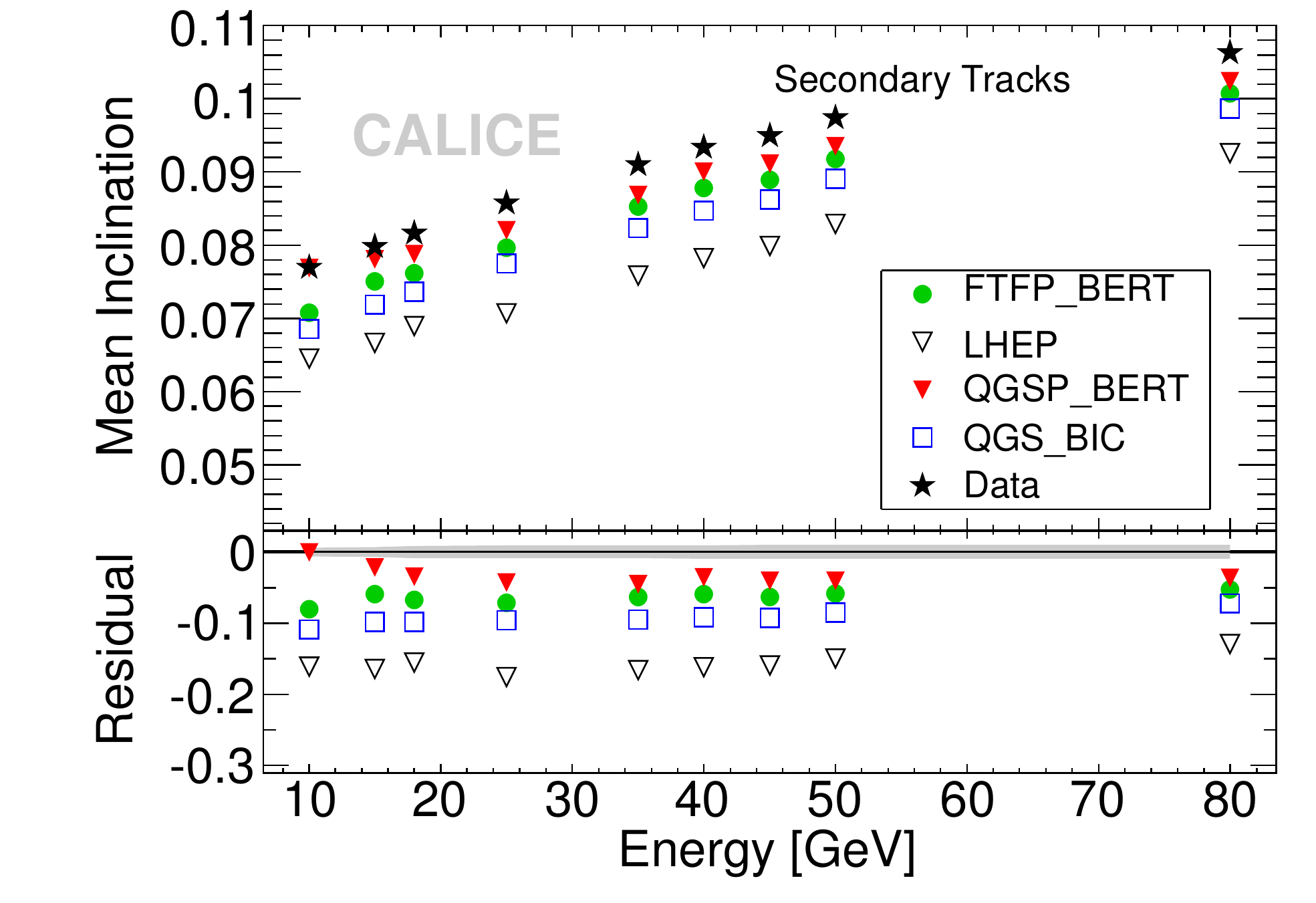}
  \caption{Mean secondary track inclination (tracks starting in layer three or later) as a function of energy. The upper panel shows the
  mean track inclination ($1-\cos \theta$) for all physics lists
  and test beam data. The lower panel shows the normalised residuals  (${\rm simulation}/{\rm data}-1$) of data and simulations with the grey area showing the statistical uncertainty of the residual of data and QGS\_BIC. Systematic errors are below the level of statistical errors, as discussed in Section \protect\ref{sec:Systematics}, and are not shown.}
  \label{fig:MCData:TrackCosPhi:MeanSec}
\end{center}
\end{figure}

Restricting the analysis to secondary tracks reduces the predominance of tracks along the beam axis, thus increasing the mean track inclination as shown in Figure \ref{fig:MCData:TrackCosPhi:MeanSec}. Also the secondary sample is still dominated by tracks at small inclinations since the production of highly energetic secondaries tends to be forward, resulting in a relatively small mean also for secondary tracks only. As in the case of the track multiplicity, the comparison to simulations gives results which are consistent with the observations for the inclusive track samples, with in general slightly smaller differences observed between data and the different physics lists.

\subsection{Track length}

The probability of a hadron to undergo an inelastic hadronic interaction with matter increases exponentially with the distance travelled. The negative inverse of the slope of this exponential distribution is given by the nuclear interaction length $\lambda_I$. The length distribution of the track segments identified by the tracking algorithm is expected to follow a similar distribution. In practice, however, the tracks are typically shorter since the tracking algorithm is not capable of identifying tracks in regions of dense shower activity and thus at the creation point of many secondary tracks. The finite single-hit efficiency can lead to a reduction of the overall length of an identified track and the presence of noise and other energy deposits can result in an early termination or the splitting of a track. In addition to a sensitivity to the quality of the description of higher-energy cross sections in the physics lists, the comparison of the track length observed in data with that in simulations provides information on the quality of the overall description of detector effects in the simulations. 

\begin{figure}
\begin{center}
  \includegraphics[width=.80\linewidth]{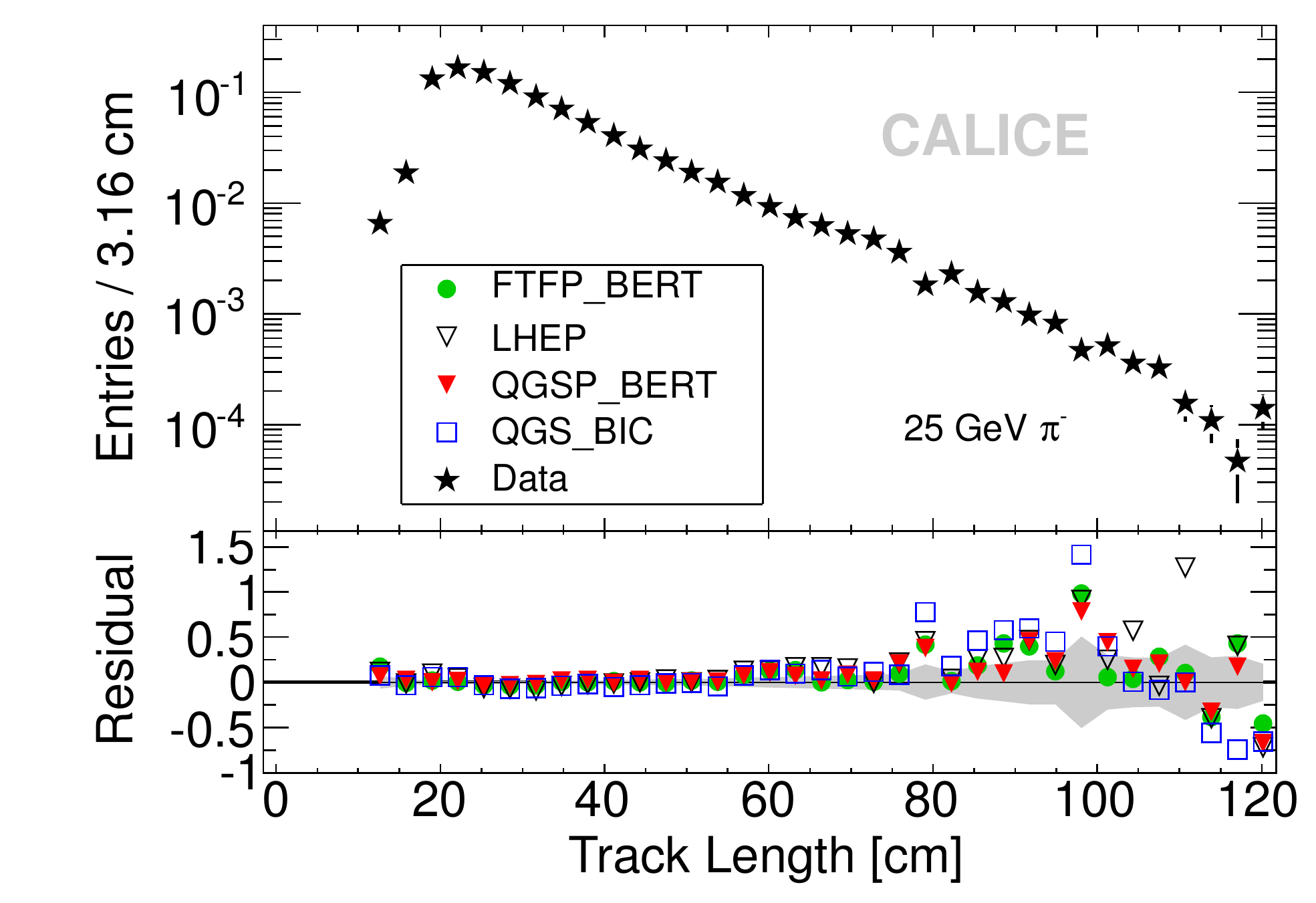}
  \caption{Normalised distribution of the length of identified tracks for a beam energy of \unit[25]{GeV}. The upper
  panel shows the distribution for data, while the lower panel shows the normalised residuals  (${\rm simulation}/{\rm data}-1$) of data and simulations. The grey area indicates the statistical uncertainty of the residual of
  QGS\_BIC compared to data.}
  \label{fig:MCData:TrackLengthSlope:25GeV}
\end{center}
\end{figure}

Figure \ref{fig:MCData:TrackLengthSlope:25GeV} shows the distribution of the length of identified tracks for a beam energy of \unit[25]{GeV} together with a comparison to simulations. The distributions are normalised to show differences in the shape of the distribution rather than in the overall number of identified tracks. All physics lists reproduce the general shape well, with larger deviations around a length of \unit[80]{cm} and around \unit[95]{cm}. These are due to reduced numbers of tracks found in data with a shower start in two layers in the front part of the calorimeter, which results in fewer tracks with a length given by the distance between these layers and the last calorimeter layer, as well as due to reduced tracking efficiency in data compared to simulations in the transition region from the fine to the coarse inner region of the detector layers. This points to an incomplete modelling of dead or noisy cells in these layers in simulations, which however does not have a significant impact on the overall results since the absolute number of affected tracks is very small.

\begin{figure}[htp]
\begin{center}
  \includegraphics[width=.80\linewidth]{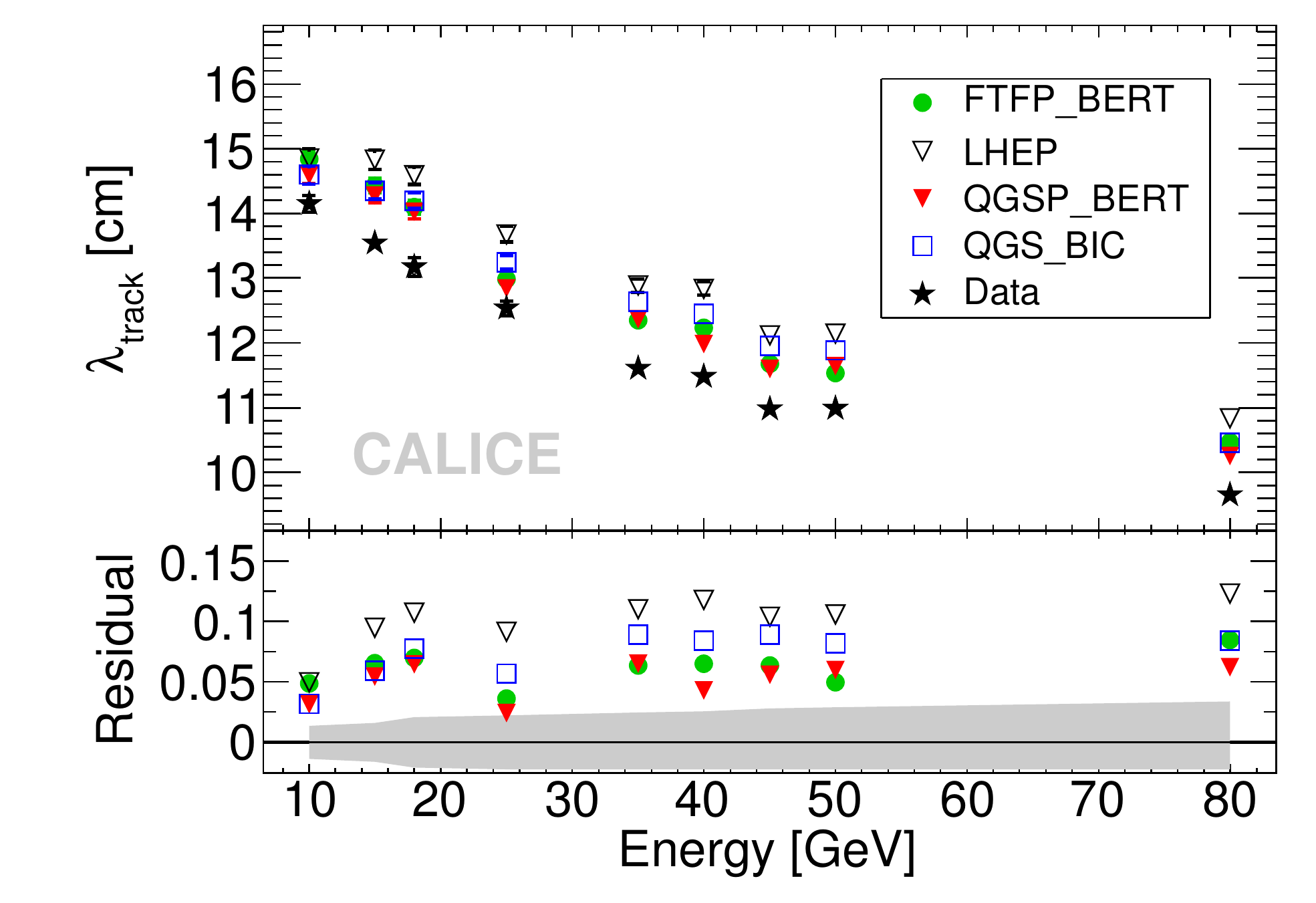}
  \caption{Negative inverse slope parameter of the exponential fit to the track length distribution as a function of beam energy. The upper panel shows data and simulations. The lower panel shows the residuals  (${\rm simulation}/{\rm data}-1$) of test beam data and simulations with the different physics lists with the
  grey band indicating the statistical uncertainties of the residuals of data and QGS\_BIC.}
  \label{fig:MCData:TrackLengthSlope:Mean}
\end{center}
\end{figure}

To study trends with beam energy, the inverse slope parameter $\lambda_{\rm track}$  is determined with an exponential fit in the range from  \unit[30]{cm} to \unit[70]{cm}, excluding problematic regions. Due to the reasons outlined above, this parameter is not identical to the nuclear interaction length of the calorimeter, but is related to it.  Figure \ref{fig:MCData:TrackLengthSlope:Mean} shows the negative inverse slope parameter $\lambda_{\rm track}$ as a function of energy for data and simulations. It decreases with increasing energy approximately following a $1/E$ behavior.  This reduction of the typical track length with increasing energy can be understood by the increase in shower activity, which increases the probability that portions of tracks are not found due to other energy deposits in the vicinity, and by the increase of the contribution of secondary tracks which originate in regions of high energy density. All physics lists studied reproduce the general trend with energy, but consistently overestimate $\lambda_{\rm track}$ and with that the track length. While QGSP\_BERT is closest to data with a typical deviation of 5\%, LHEP overestimates the slope parameter by 10\%. The observed differences point to differences in shower shape and an underestimate of spurious hits which can lead to a truncation of tracks, but can also partially originate from the treatment of noise hits in simulations as discussed below.

\begin{figure}[htp]
\begin{center}
  \includegraphics[width=.80\linewidth]{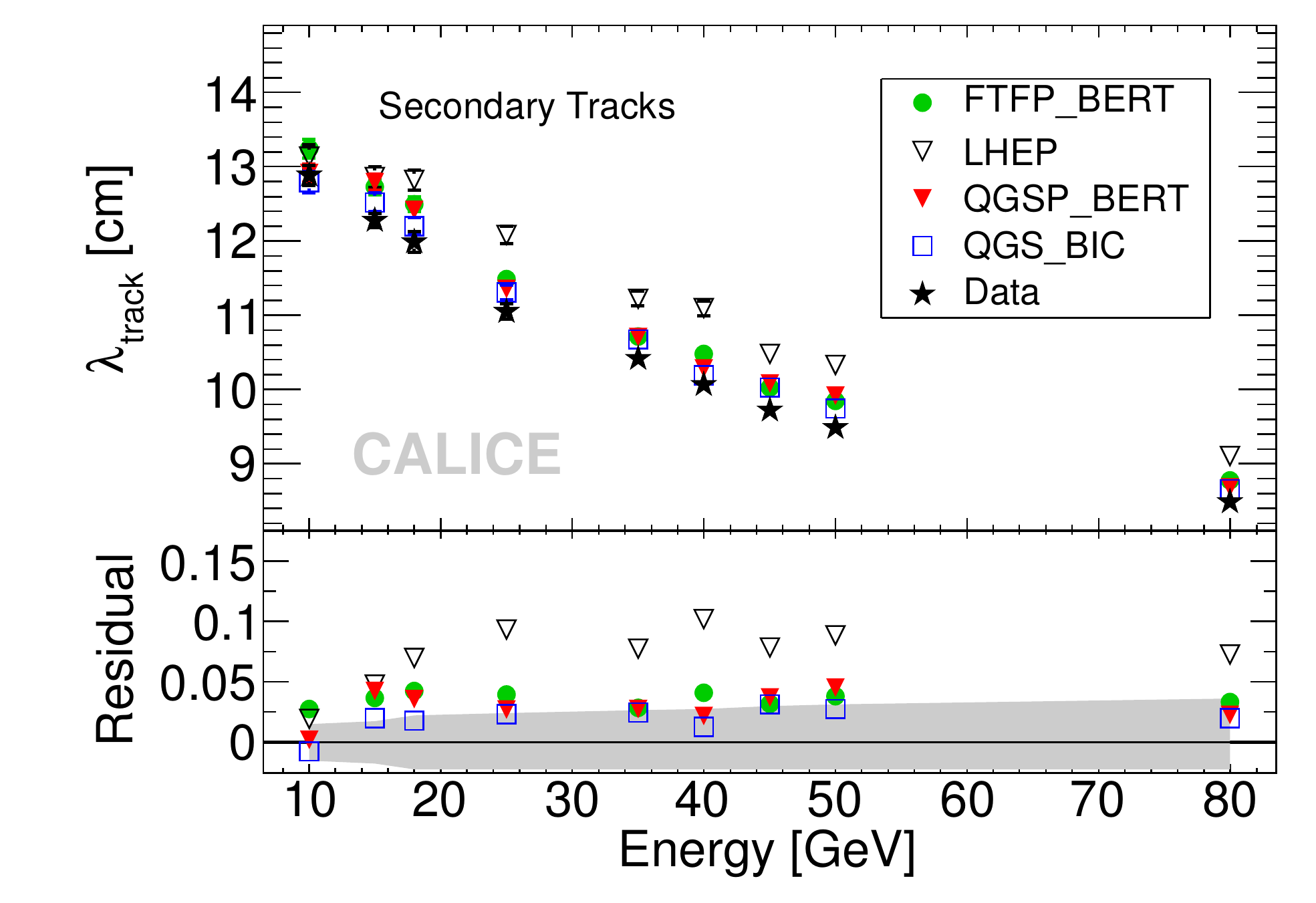}
  \caption{Negative inverse slope parameter of the exponential fit to the track length distribution as a function of beam energy for secondary tracks (starting in layer three or later). The upper panel shows data and simulations. The lower panel shows the residuals  (${\rm simulation}/{\rm data}-1$) of test beam data and simulations with the different physics lists with the
  grey band indicating the statistical uncertainties of the residuals of data and QGS\_BIC.}
  \label{fig:MCData:TrackLengthSlope:MeanSec}
\end{center}
\end{figure}

The negative inverse slope parameter for secondary tracks, shown in Figure \ref{fig:MCData:TrackLengthSlope:MeanSec} is lower than the one observed for primary tracks, as expected from the bigger impact of the regions with higher shower density in the vicinity of inelastic interactions. The general trend with energy is identical to the one observed for the inclusive sample shown in Figure \ref{fig:MCData:TrackLengthSlope:Mean}. As for the other observables as well, the comparison to simulations for secondary tracks only yields results which are consistent with the observations for all tracks. In general the difference between data and simulations are somewhat smaller for the secondary tracks, with the largest deviations observed for LHEP, which overestimates the slope parameter by approximately 8\%.

\subsection{Systematic Uncertainties}
\label{sec:Systematics}

Systematic uncertainties on the track finding performance could potentially affect the conclusions drawn from the comparison of data and simulations. The sizes of possible effects are studied here using simulations with the QGSP\_BERT physics list by comparing the results obtained with the standard simulations with those with modified parameters. The identification of track segments is a rather robust technique since it is purely based on the identification of hits irrespective of their precise energy content, and as such does not impose strict requirements on the control of the energy calibration of the detector. Nevertheless, there are two potential sources of systematic effects which are studied in the following, namely the MIP energy scale given by the precision of the single cell calibration and the modelling of the detector noise.

\subsubsection{MIP energy scale}

Each cell of the AHCAL is calibrated with muons recorded in dedicated calibration runs. The most probable value of the response to the penetrating MIPs is used as the cell-to-cell calibration scale and as the reference energy scale in the tracking algorithm, as discussed in Section \ref{sec:EventSelection}. The exact calibration procedure is described in \cite{Adloff:2010hb}. The uncertainty of these calibration factors is approximately 2\%, originating from statistical and fit systematic uncertainties \cite{collaboration:2010rq}. 

In the present analysis, the cell energy information is only used for noise rejection by requiring a minimum energy of  \unit[0.5]{MIP}. The uncertainty of the calibration can therefore have an influence on the single cell efficiency and thus on the track finding performance. The influence of an uncertainty of the MIP energy scale is studied by altering the threshold, which provides an upper limit on possible effects since this corresponds to a correlated shift of the calibration of all detector cells. In addition, rather large shifts of the threshold of $\pm4\%$ and $\pm8\%$, corresponding to values of \unit[0.46, 0.48, 0.52 and 0.54]{MIPs} respectively, are studied. Even these very conservative values have little influence on all observables studied in the present article. With threshold shifts of  $\pm8\%$ the changes are comparable to or slightly in excess of the statistical error, while smaller shifts result in variations significantly below the statistical uncertainties. Overall, systematic effects originating from the MIP energy scale uncertainty are negligible, with respect to the data-simulation comparisons.

\subsubsection{Detector Noise}

\begin{figure}
\begin{center}
  \includegraphics[width=.80\linewidth]{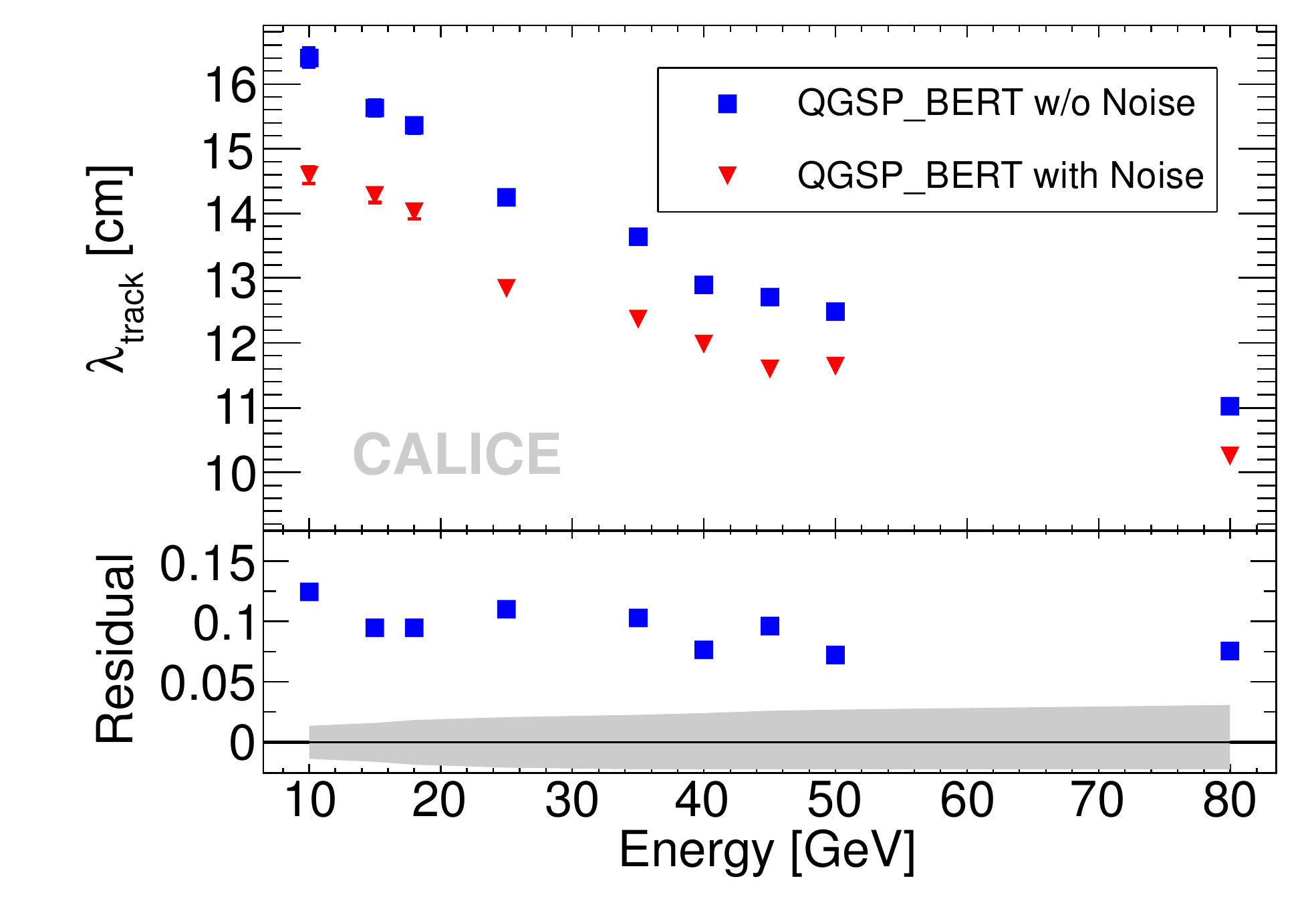}
  \caption{Comparison of the inverse slope parameter of the track length distribution as a function of energy for simulations with and without the inclusion of detector noise.
  The residual is defined here as $\frac{\textrm{QGSP\_BERT w/o Noise}}{\textrm{QGSP\_BERT with Noise}}-1$, and its statistical uncertainty is indicated by the grey band.}
  \label{fig:Noise}
\end{center}
\end{figure}

The detector noise has a significant impact on the performance of the tracking algorithm, since only isolated hits are considered, irrespective of their energy content. On one hand, the addition of noise hits, which often are isolated since they occur at random locations inside the detector, can lead to an increase in track length from the addition of noise hits to real tracks. On the other hand, additional hits also lead to a reduction of the number of real isolated hits which are used in the tracking, and thus reduce the length of identified tracks. 
In practice, this second effect far outweighs the first. The influence of noise on the track segments found is studied by comparing results from full simulations with simulations where no detector noise from data has been overlaid. While no significant effect on the track multiplicity and on the angular distribution is observed the inverse slope parameter $\lambda_{\rm track}$ of the track length distribution decreases by 5\% to 10\% once detector noise is included, as shown in  Figure\ \ref{fig:Noise}. The influence of noise on the track length is comparable to the difference observed between data and simulations. Since simulations use detector noise taken directly from the test beam data they are compared to, the uncertainties on the noise level in the simulations are significantly smaller than the noise level itself, and cannot alone explain the discrepancy between data and simulations.

\subsection{Track segments as a calibration tool}

\begin{figure}
\begin{center}
  \includegraphics[width=.49\linewidth]{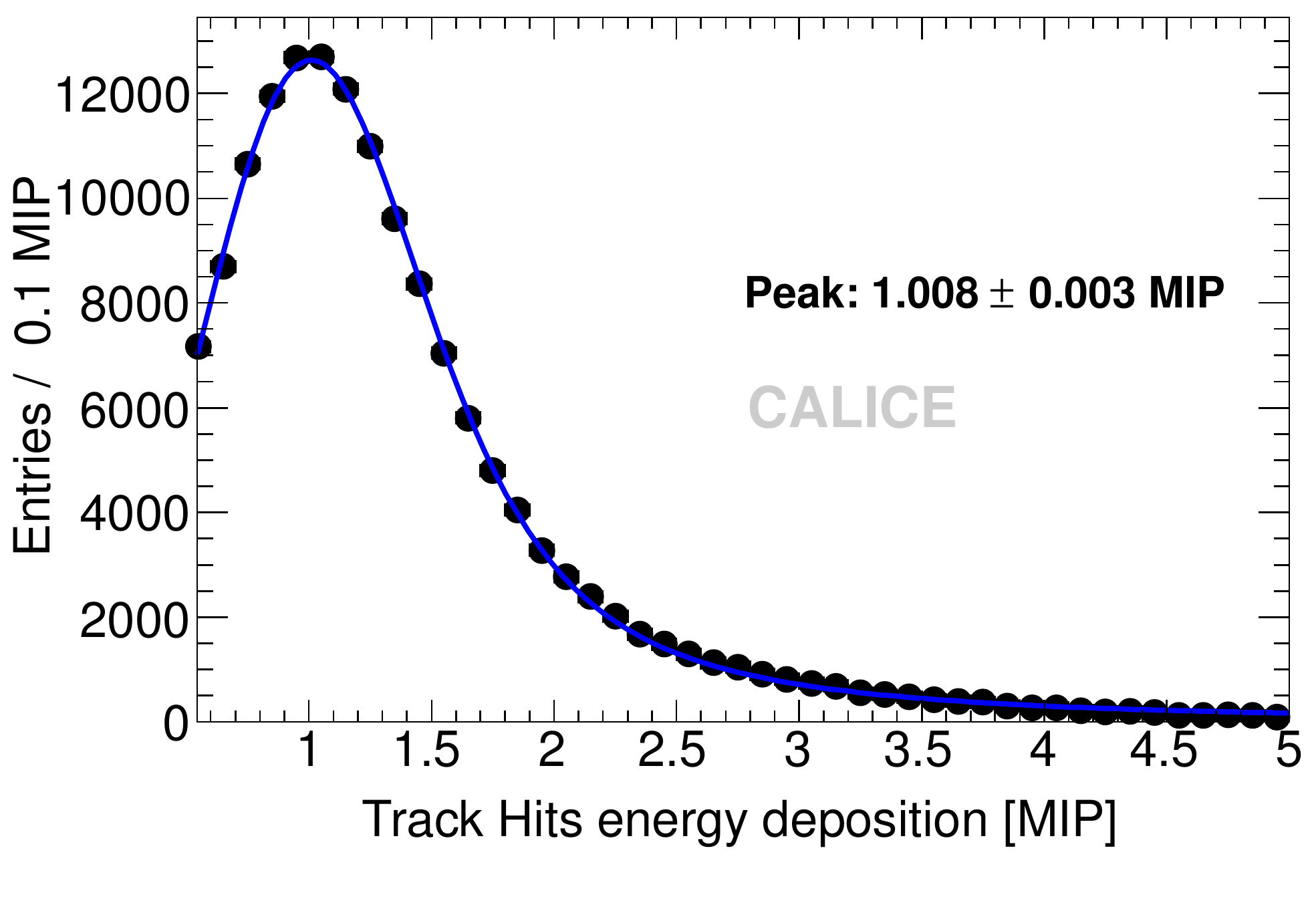}%
  \includegraphics[width=.49\linewidth]{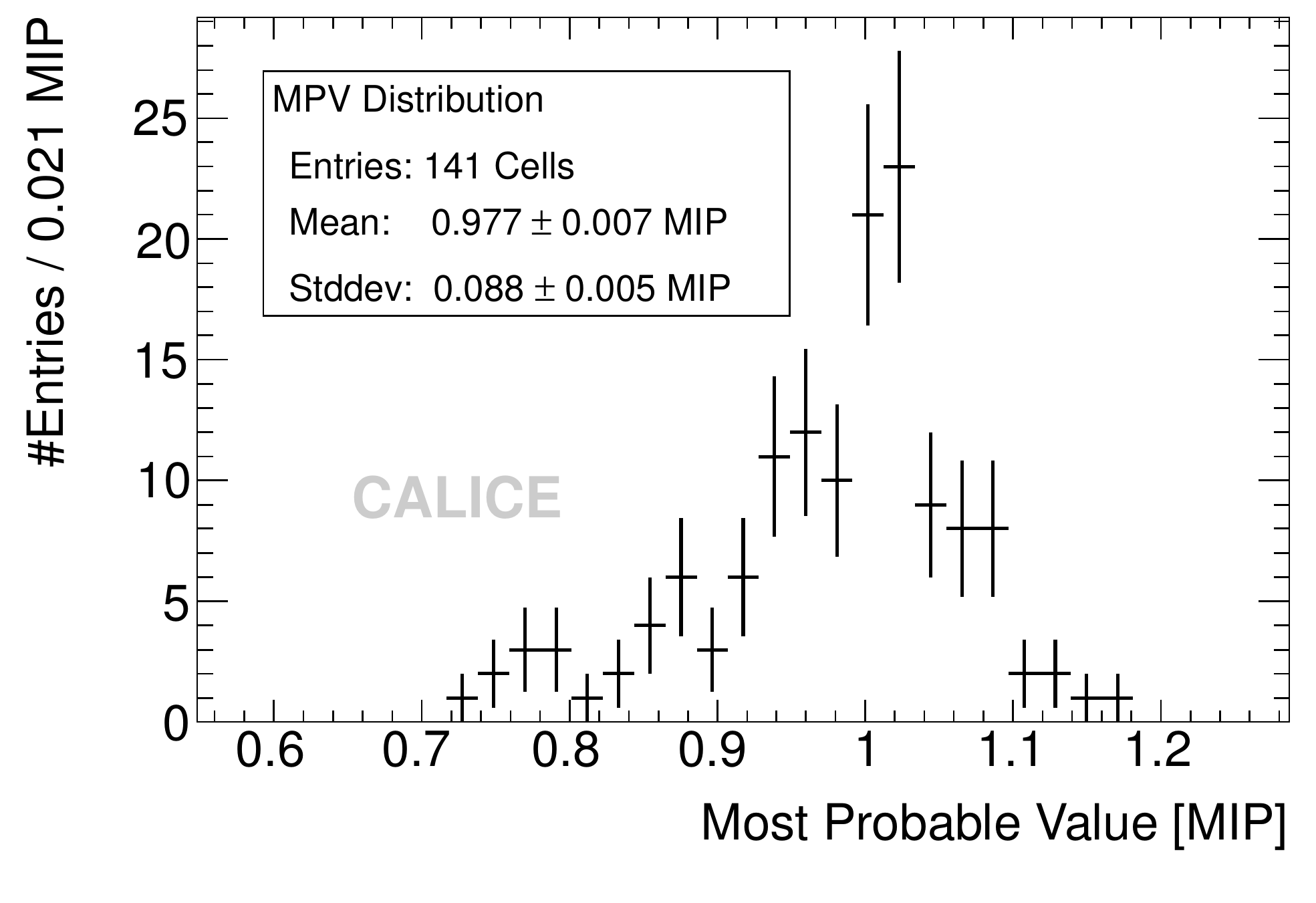}
  \caption{
  {\it Left:} Distribution of cell signal amplitude on identified tracks in \unit[80]{GeV} pion showers with small angles with respect to the beam axis, satisfying  cos$\theta > 0.98$. The distribution is fit with a convolution of a Landau function and a Gaussian to describe the energy deposition together with the response of the photon sensor and the readout electronics.
  {\it Right:} Most probable values for all 141 cells with at least 100 entries in the \unit[80]{GeV} run, with the mean and the standard deviation of the distribution given for illustration. The width of the distribution is dominated by the statistical errors of the individual fits. 
  }
  \label{fig:Edep}
\end{center}
\end{figure}

In addition to the possibility of probing the three-dimensional structure of hadronic cascades, min\-i\-mum-ionizing track segments identified in hadronic showers can also serve as calibration tools to monitor the single cell response of a highly granular calorimeter {\it in situ} without  dedicated calibration runs. Figure \ref{fig:Edep} {\it left} shows the data distribution of the single cell amplitude in identified track segments in \unit[80]{GeV} pion showers. To avoid corrections for the track length in the cell due to larger angles with respect to the beam axis, only tracks at small angles, satisfying cos$\theta > 0.98$ are considered. This sample contains tracks from particles prior (i.e.~primary tracks) and post (i.e.~secondary tracks) the first hard interaction, but is dominated by the former. 
The distribution is fit with a convolution of a Landau function with a Gaussian, where the former accounts for the energy loss of minimum-ionising particles in the scintillator tiles and the latter for effects from photon statistics, SiPM response and contributions from the detector front-end electronics. The most probable value of the distribution is consistent with \unit[1]{MIP}, the calibration factor set by calibration runs with muons. This good agreement demonstrates the performance of the corrections applied to account for the dependence of the detector response on temperature and operational parameters, and shows that the identified track segments are indeed minimum-ionising particles. 
The cell-to-cell variations of the distribution are displayed in Figure \ref{fig:Edep} {\it right} by showing the distribution of the measured most probable value for all cells with at least 100 entries in the studied \unit[80]{GeV} data set where the error on the most probable value was below \unit[0.3]{MIP}. The width of the distribution is dominated by the fit uncertainty of the most probable value and not by cell-to-cell response variations. The distribution of fit errors has a mean value of 4.2\% and extends substantially beyond 10\% for some of the cells with low statistics. Overall, these results demonstrate that track segments identified within hadronic showers can be used to calibrate and monitor highly granular calorimeter systems at future colliders. To increase the statistics for such a monitoring algorithm, also non-isolated hits could be considered, identified as belonging to a track by interpolating between already identified track segments.

\section{Summary}
\label{sec:Summary}

We have studied the spatial structure of hadronic showers in a highly granular scintillator-steel sampling calorimeter by identifying track segments of minimum-ionising particles within showers induced by negative pions with an energy from \unit[10]{GeV} to \unit[80]{GeV}. The tracks are identified with a nearest neighbour algorithm considering isolated hits, providing sensitivity to the outer regions of hadronic showers. This analysis demonstrates the imaging capabilities of the CALICE AHCAL, which is also crucial for the performance of particle flow algorithms. The single-cell energies of hits on identified tracks show a distribution expected for minimum-ionising particles, making such tracks within hadronic showers suitable for an in-situ cell-to-cell calibration of highly granular calorimeters at future colliders. In {\sc Geant}4 simulations the reconstructed track multiplicity is correlated with the total number of charged higher-energetic particles (excluding electrons and positrons), providing the basis for tests of {\sc Geant}4 hadronic shower models with observables based on the reconstructed tracks. Four physics lists of {\sc Geant}4 9.4p02 have been investigated. The observations for all identified track segments and those for secondary tracks only are consistent with each other. QGSP\_BERT generally provides the best agreement with data with a good reproduction of the track multiplicity, while LHEP shows the largest discrepancies, producing too few secondary tracks overall and strongly underestimating the number of tracks at large angles.

\section{Acknowledgements}

We gratefully acknowledge the DESY and CERN managements for their support and
hospitality, and their accelerator staff for the reliable and efficient
beam operation. 
We would like to thank the HEP group of the University of Tsukuba for the loan
of drift chambers for the DESY test beam. 
The authors would like to thank the RIMST (Zelenograd) group for their
help and sensors manufacturing. 
This work was supported by the Bundesministerium f\"{u}r Bildung und Forschung, Germany; 
by the  the DFG cluster of excellence `Origin and Structure of the Universe' of Germany;  
by the Helmholtz-Nachwuchsgruppen grant VH-NG-206; 
by the BMBF, grant no. 05HS6VH1; 
by the Alexander von Humboldt Foundation (Research Award IV, RUS1066839 GSA); 
by joint Helmholtz Foundation and RFBR grant HRJRG-002, SC Rosatom; 
by the Russian Ministry of Education and Science via
grants 8174, 8411, 1366.2012.2, P220;
by MICINN and CPAN, Spain;
by CRI(MST) of MOST/KOSEF in Korea; 
by the US Department of Energy and the US National Science Foundation; 
by the Ministry of Education, Youth and Sports of the Czech Republic
under the projects AV0 Z3407391, AV0 Z10100502, LC527  and LA09042 and by the
Grant Agency of the Czech Republic under the project 202/05/0653;  
by the National Sciences and Engineering Research Council of Canada; 
and by the Science and Technology Facilities Council, UK.

\bibliography{bibliography}

\providecommand{\href}[2]{#2}\begingroup\raggedright\begin{thebibliography}{10}

\bibitem{Thomson:2009rp}
M.~Thomson, {\it {Particle Flow Calorimetry and the PandoraPFA Algorithm}},
  {\em Nucl. Instrum. Meth.} {\bf A611} (2009) 25--40,
  [\href{http://xxx.lanl.gov/abs/0907.3577}{{\tt arXiv:0907.3577}}].

\bibitem{Brient:2002gh}
J.-C. Brient and H.~Videau, {\it {The Calorimetry at the future $e^+e^-$ linear
  collider}},  {\em eConf} {\bf C010630} (2001) E3047,
  [\href{http://xxx.lanl.gov/abs/hep-ex/0202004}{{\tt hep-ex/0202004}}].

\bibitem{pfaMorgunov}
V.~L. Morgunov, {\it {Calorimetry design with energy-flow concept (imaging
  detector for high-energy physics)}},  {\em {CALOR 2002, Pasadena, California.
  Published in Pasadena 2002, 'Calorimetry in particle physics'}} (2002).

\bibitem{Adloff:2010hb}
{CALICE} Collaboration, C.~Adloff {\em et.~al.}, {\it {Construction and
  Commissioning of the CALICE Analog Hadron Calorimeter Prototype}},  {\em
  JINST} {\bf 5} (2010) P05004, [\href{http://xxx.lanl.gov/abs/1003.2662}{{\tt
  arXiv:1003.2662}}].

\bibitem{Bondarenko:2000in}
G.~Bondarenko, P.~Buzhan, B.~Dolgoshein, V.~Golovin, E.~Gushchin, {\em
  et.~al.}, {\it {Limited Geiger-mode microcell silicon photodiode: New
  results}},  {\em Nucl. Instrum. Meth.} {\bf A442} (2000) 187--192.

\bibitem{Buzhan:2003ur}
P.~Buzhan, B.~Dolgoshein, L.~Filatov, A.~Ilyin, V.~Kantserov, {\em et.~al.},
  {\it {Silicon photomultiplier and its possible applications}},  {\em Nucl.
  Instrum. Meth.} {\bf A504} (2003) 48--52.

\bibitem{Anduze:2008hq}
{CALICE} Collaboration, J.~Repond {\em et.~al.}, {\it {Design and Electronics
  Commissioning of the Physics Prototype of a Si-W Electromagnetic Calorimeter
  for the International Linear Collider}},  {\em JINST} {\bf 3} (2008) P08001,
  [\href{http://xxx.lanl.gov/abs/0805.4833}{{\tt arXiv:0805.4833}}].

\bibitem{CALICE:2012aa}
{CALICE} Collaboration, C.~Adloff {\em et.~al.}, {\it {Construction and
  performance of a silicon photomultiplier/extruded scintillator tail-catcher
  and muon-tracker}},  {\em JINST} {\bf 7} (2012) P04015,
  [\href{http://xxx.lanl.gov/abs/1201.1653}{{\tt arXiv:1201.1653}}].

\bibitem{Agostinelli:2002hh}
{GEANT4} Collaboration, S.~Agostinelli {\em et.~al.}, {\it {GEANT4: A
  Simulation toolkit}},  {\em Nucl. Instrum. Meth.} {\bf A506} (2003) 250--303.

\bibitem{Fruehwirt:DataAnalysis:Hough}
R.~Fr\"{u}wirth, M.~Regler, R.~K. Bock, H.~Grote, and D.~Notz, {\em {Data
  Analysis Techniques for High-Energy Physics}}.
\newblock Cambridge monographs on Particle Physics, Nuclear Physics and
  Cosmology, second~ed., 2000.

\bibitem{collaboration:2010rq}
{CALICE} Collaboration, C.~Adloff {\em et.~al.}, {\it {Electromagnetic response
  of a highly granular hadronic calorimeter}},  {\em JINST} {\bf 6} (2011)
  P04003, [\href{http://xxx.lanl.gov/abs/1012.4343}{{\tt arXiv:1012.4343}}].

\bibitem{Geant4:PhysicsLists}
A.~Ribon, J.~Apostolakis, A.~Dotti, G.~Folger, V.~Ivanchenko, M.~Kosov,
  V.~Uzhinsky, and D.~H. Wright, {\it {Status of Geant4 hadronic physics for
  the simulation of LHC experiments at the start of LHC physics program}},
  {\em CERN-LCGAPP-2010-02} (2010).

\bibitem{Geant4:PhysicsLists:Improvements}
A.~Dotti, J.~Apostolakis, G.~Folger, V.~Grichine, V.~Ivanchenko, {\em et.~al.},
  {\it {Recent improvements on the description of hadronic interactions in
  Geant4}},  {\em J. Phys. Conf. Ser.} {\bf 293} (2011) 012022.

\bibitem{Fesefeldt:1985yw}
H.~Fesefeldt, {\it {The simulation of hadronic showers: physics and
  applications}},  {\em PITHA-85-02} (1985).

\bibitem{Folger:2003sb}
G.~Folger and J.~Wellisch, {\it {String parton models in GEANT4}},  {\em eConf}
  {\bf C0303241} (2003) MOMT007,
  [\href{http://xxx.lanl.gov/abs/nucl-th/0306007}{{\tt nucl-th/0306007}}].

\bibitem{Heikkinen:2003sc}
A.~Heikkinen, N.~Stepanov, and J.~P. Wellisch, {\it {Bertini intranuclear
  cascade implementation in GEANT4}},  {\em eConf} {\bf C0303241} (2003)
  MOMT008, [\href{http://xxx.lanl.gov/abs/nucl-th/0306008}{{\tt
  nucl-th/0306008}}].

\bibitem{NilssonAlmqvist:1986rx}
B.~Nilsson-Almqvist and E.~Stenlund, {\it {Interactions Between Hadrons and
  Nuclei: The Lund Monte Carlo, Fritiof Version 1.6}},  {\em Comput. Phys.
  Commun.} {\bf 43} (1987) 387.

\bibitem{BIC}
G.~Folger, V.~N. Ivanchenko, and J.~P. Wellisch, {\it The binary cascade},
  {\em Eur. Phys. J} {\bf A21} (2004) 407.

\end{thebibliography}\endgroup

\end{document}